\documentclass[final,5p,times,twocolumn,authoryear]{elsarticle}
\usepackage[utf8]{inputenc}
\usepackage{amsmath}
\usepackage{amsfonts}
\usepackage{amssymb}
\usepackage{graphicx}
\usepackage{subcaption}
\usepackage{array}
\usepackage{afterpage}
\usepackage{placeins}
\usepackage{comment}
\usepackage{enumitem}
\usepackage{tikz}
\usepackage{xcolor}
\usepackage{chngcntr}
\usepackage{float}
\usepackage{url}
\usepackage{multirow}
\journal{Astronomy $\&$ Computing}
\usepackage{tabularx}
\usepackage[colorlinks=true,linkcolor=blue]{hyperref}

\begin{document}
\begin{frontmatter}
\title{DarsakX: A Python Package for Designing and Analyzing Imaging Performance of X-ray Telescopes}

\author[prl,iitgn]{Neeraj K. Tiwari}
\author[prl]{ Santosh V. Vadawale}
\author[prl]{N. P. S. Mithun}
\author[prl]{C. S. Vaishnava}
\author[prl]{ Bharath Saiguhan}
\affiliation[prl]{organization={Physical Research Laboratory},
            city={Ahmedabad},
            postcode={380009},
            state={Gujarat},
            country={India}}
            
\affiliation[iitgn]{organization={Indian Institute of Technology Gandhinagar},
            city={Gandhinagar},
            postcode={382355},
            state={Gujarat},
            country={India}}
\begin{abstract}

The imaging performance and sensitivity of an X-ray telescope when observing astrophysical sources are primarily governed by the optical design, geometrical uncertainties (figure errors, surface roughness, and mirror alignment inaccuracies), and the reflectivity properties of the X-ray reflecting mirror surface. To thoroughly evaluate the imaging performance of an X-ray telescope with an optical design similar to Wolter-1 optics, which comprises multiple shells with known geometrical uncertainties and mirror reflectivity properties, appropriate computational tools are essential. These tools are used to estimate the angular resolution and effective area for various source energies and locations and, more importantly, to assess the impact of figure errors on the telescope's imaging performance. Additionally, they can also be used to optimize optics geometry by modifying it in reference to the Wolter-1 optics, aiming to minimize the optical aberration associated with the Wolter-1 configuration. In this paper, we introduce DarsakX, a Python-based ray tracing computational tool specifically designed to estimate the imaging performance of a multi-shell X-ray telescope. DarsakX has the capability to simulate the impact of figure errors present in the axial direction of a mirror shell. The geometrical shape of the mirror shells can be defined as a combination of figure error with the base optics, such as Wolter-1 or Conical optics. Additionally, DarsakX allows the exploration of new optical designs involving two reflections similar to Wolter-1 optics but with an improved angular resolution for wide-field telescopes. Developed through an analytical approach, DarsakX ensures computational efficiency, enabling fast processing.
\end{abstract}
\begin{keyword}
X-ray Astronomy \sep X-ray Optics \sep Ray-Tracing \sep Wolter Optics
\end{keyword}
\end{frontmatter}

\section{Introduction}
X-ray telescopes are crucial for exploring the X-ray universe as they enhance observational sensitivity and spatial resolution to resolve nearby sources. When observing extended sources such as supernovas, galaxies, and galaxy clusters, these telescopes can provide detailed images and enable high-sensitivity studies of their local regions. While surveying the sky or observing the Sun in X-rays, imaging telescopes with a wide Field of View (FOV$\approx1^\circ$) become essential. However, achieving uniform performance across a wide FOV remains a technological challenge \citep{{conconi2010wide},{rosati2010wide}}. The ultimate performance of a telescope is determined by evaluating its effective area and angular resolution as functions of source energy and location. The primary objective in advancing X-ray telescopes is to improve both effective area and angular resolution uniformly across a wide FOV and energy. 

Astronomical X-ray telescopes primarily function based on the principle of reflection at grazing incident angles. The X-ray reflecting surface must be exceptionally smooth, with an RMS value comparable to the X-ray wavelength for specular reflection \citep{aschenbach1985x}. To achieve total external reflection, it requires a highly grazing incident angle ($\textless1^\circ$, where the ray is almost parallel to the reflecting surface) and a surface composed of high atomic number (Z) elements, especially for soft X-rays. For hard X-rays, the surface needs to be constructed with a multilayer structure comprising alternating high and low Z elements \citep{mondal2021darpanx,windt2015advancements}. 

X-ray telescopes primarily rely on the Wolter-1 optics ($WO_1$), utilizing dual reflections to minimize optical aberrations and enhance off-axis angular resolution \citep{wolter1952spiegelsysteme}. While the mirror's figure accuracy influences on-axis angular resolution \citep{wu2022influence}, off-axis performance is constrained by inherent optical aberrations linked to the $WO_1$ design itself \citep{{shealy1990formula},{vanspeybroeck1972design}}. Due to a requirement for a grazing incident angle, the geometrical area of the $WO_1$ optics is notably small. Therefore, to enhance the effective area of the telescope, multiple coaxially aligned mirror shells in the form of $WO_1$ optics are used, all sharing a common focus.

The effective area and angular resolution of X-ray telescopes have improved gradually over the past four decades with the launch of various telescopes such as Einstein \citep{giacconi1979einstein}, EXOSAT \citep{de1981x}, ROSAT \citep{trumper1982rosat}, ASCA \citep{tanaka1994x}, Chandra \citep{weisskopf2002overview}, XMM-Newton \citep{jansen2001xmm}, Swift XRT \citep{burrows2005swift}, NuSTAR \citep{harrison2013nuclear}, eROSITA \citep{predehl2021erosita}, and XRISM \citep{tashiro2022xrism}. Each telescope was optimized for specific objectives, aiming to achieve different angular resolutions, effective areas, FOV, and energy ranges.

The next generation of X-ray telescopes requires further enhancements in effective area and angular resolution to advance our understanding of X-ray-emitting celestial sources. Both aspects are essential for improving the telescope's sensitivity to faint signals \citep{o2010high}. However, high angular resolution becomes even more crucial when imaging extended sources, enabling spectroscopy, time variability, and polarization studies of the local region of the extended source. Simultaneously achieving high angular resolution and effective area poses a challenge \citep{o2010high}. For instance, the Chandra telescope is constructed with high angular resolution but moderate effective area, whereas the XMM-Newton telescope is constructed with the opposite characteristic. Additionally, ensuring uniformity of these aspects across the entire FOV and energy band presents technical challenges. For example, Chandra's angular resolution degrades from 0.5 arcsec at the on-axis source to 13 arcsec at a field angle of 15 arcmin\footnote{https://cxc.harvard.edu/proposer/POG/}. At the same time, its effective area decreases from 785~$\mathrm{{cm}^2}$ at 1~$\mathrm{keV}$ to 33~$\mathrm{{cm}^2}$ at 10~$\mathrm{keV}$ \citep{zhao2004chandra}.

The on-axis angular resolution of an X-ray telescope, mainly affected by geometrical errors, can be improved by reducing these errors. The geometrical errors in the telescope mainly result from fabrication processes and assembly methods, allowing improvement through an iterative process. A ray tracing simulation tool is essential to assess a telescope's imaging performance with given geometrical errors and mirror coating parameters. These assessments help in estimating the impact of fabrication techniques and in refining them to optimize the overall telescope performance.

Ray-tracing tools are also crucial to optimize telescope performance for a wide FOV, such as observing the Sun or extended astrophysical sources, where a direct $WO_1$ configuration does not provide very good angular resolution. Altering the figure surface of $WO_1$ by incorporating additional figure surface can enhance off-axis angular resolution \citep{{conconi2010wide},{saha2022optical},{werner1977imaging}}. The Wolter–Schwarzschild (WS) and Hyperboloid–Hyperboloid (HH) designs may provide better angular resolution than $WO_1$ for off-axis sources  \citep{{wolter1952verallgemeinerte},{chase1973wolter},{harvey2001grazing},{saha2014optical}}. However, the deviation in the figure of WS, and HH designs from the $WO_1$ configuration is on the order of microns. Therefore, a ray tracing tool capable of accommodating micron-level figure errors can also be utilized to optimize the wide FOV telescope design by altering the $WO_1$ figure. Additionally, these tools are crucial for analyzing observational data and mapping source images from the sky to the detector plane and vice versa.

Various X-ray observatories have developed ray-tracing tools to design their optics and analyze observational data. For instance, Chandra employs the ray-tracing tool ChaRT (web-based) and SAOTrace (system-based), both dedicated exclusively to Chandra optics \citep{jerius2004role}. These tools also consider the reflectivity properties of the mirrors and account for geometrical errors by incorporating metrology data. Similarly, TraceIT has been utilized for XMM-Newton \citep{sironi2011x}, and MT-RAYOR has been employed for NuSTAR optics \citep{{westergaard2011mt_rayor},{westergaard2012nustar}}.

In this paper, we introduce a ray-tracing tool named DarsakX (`Darsak' meaning `observer' in Sanskrit), designed to assess the imaging performance of X-ray telescopes, as part of the ongoing efforts towards the future Indian X-ray observatory program. Previously, we developed a tool named DarpanX \citep{mondal2021darpanx} to evaluate the reflectivity properties of mirror shells based on given surface parameters. These reflectivity properties, estimated by DarpanX, are utilized within DarsakX to compute parameters such as the effective area and angular resolution of the telescope.

DarsakX has been developed as an open-source Python package designed to trace rays in 3D for multi-shell X-ray telescopes. It has the ability to consider figure errors present in the axial direction. Additionally, DarsakX can optimize the geometrical shape of the optics, allowing for control over the angular resolution across a wide FOV. For example, it can improve the off-axis angular resolution by compromising the on-axis resolution to make it more uniform across the entire FOV, which is applicable to solar telescopes. Developed through an analytical approach, DarsakX ensures computational efficiency. When the need for a numerical solution arises, such as when tracing rays through a mirror with complex non-linear figure errors, an alternate option is provided to trace the rays analytically, based on valid approximate assumptions.  

The functional overview of DarsakX is presented in Section \ref{sec:FunctionalOverview}. The ray-tracing methodology for a multi-shell X-ray telescope is detailed in Section \ref{sec:01}. Section \ref{sec:02} presents the techniques used to assess various performance parameters of the telescope. Validation and application of DarsakX are discussed in Section \ref{sec:04}.

\section{DarsakX: An Overview}
\label{sec:FunctionalOverview}
In this section, we provide an overview of the optical configuration considered in DarsakX, as well as the ray tracing methodology and its implementation.

\subsection{Geometrical Definition}
The telescope can consist of a single shell or coaxially aligned multiple shells, where each shell comprises a combination of a base surface and a figure surface. Here, the base surface is the ideal optical surface. It can be either $WO_1$ or Conical Optics (CO, a conical approximation to $WO_1$). The figure surface represents micron-level modulation on top of the base surface, affecting the angular resolution of the mirror shell without significantly impacting its effective area. A shell comprises two axially symmetric mirrors. The primary mirror is defined by either a paraboloid surface (when the base surface is $WO_1$) combined with the figure surface or a conical surface (when the base surface is $CO$) combined with the figure surface, used for the first reflection. The secondary mirror, for the second reflection following the primary, is similarly defined, either as a hyperboloid surface (when the base surface is $WO_1$) combined with the figure surface or a conical surface (when the base surface is $CO$) combined with the figure surface. With the base surface being axially symmetric, we have assumed the figure surface to also follow this symmetry. Therefore, the shell is always axially symmetric. The figure surface serves a dual purpose; it can be utilized to define the figure error present in the base optics acquired during fabrication, contributing to a more realistic assessment of imaging performance. Alternatively, it can act as a user-defined surface to alter the geometrical optics with respect to base optics. By changing geometrical optics, angular resolution can be improved over a wide field of view.  It's essential to note that in both scenarios, the figure surface needs to be continuous and differentiable. 
When parallel rays originating from a point source at infinity pass through the telescope, they are expected to converge onto a focal point. However, due to optical aberrations, figure errors, and surface roughness, they spread across the focal plane instead of focusing precisely on a point. This distribution is defined as the point spread function (PSF). In the context of X-ray optics, PSF has two components: core and wing. The PSF core is the spread due to specular reflection, which depends on optical aberration and figure error present in the optics, and it remains independent of X-ray energy. Whereas, the wing is the radially extended region of the PSF until a large diameter compared to the core, and it is produced due to X-ray scattering caused by surface roughness in the mirror, depending on the energy of the photon.
At present, DarsakX
does not take into account the influence of surface roughness on
the shape of the PSF. The effect of surface roughness on the PSF shape can be addressed using X-ray scattering theory from rough surfaces or through the Huygens-Fresnel principle \citep{{spiga2007analytical},{raimondi2011point}}.

\subsection{Source Definition}
In DarsakX, a point source position at infinity is defined by $(\theta, \phi)$, where $\theta$ represents the angle between the source direction and the telescope's optical axis (designated as the $X$ axis in Figure \ref{fig:A1}). The angle $\phi$ denotes the angle between the source direction on the detector plane ($YZ$ plane) and the $Y$ axis. Given the telescope's assumed axial symmetry, $\phi=0$ is always considered.

\subsection{Ray Tracing Methodology}
In the DarsakX ray tracing methodology, each ray represents either a beam of photons or a single X-ray photon and is assigned an effective area. The ray’s effective area can be adjusted by altering the ray density ($\mathrm{rays/cm^2}$), where higher ray density corresponds to a lower effective area per ray. DarsakX assumes uniform ray density across the entire beam area and ensures that the beam diameter exceeds the telescope's aperture. As rays undergo two reflections with mirrors, their individual associated effective area decreases due to the mirror's reflective properties. The sum of the individual effective areas of all rays that reach the detector provides the telescope's total effective area.

In DarsakX, rays are traced through each individual shell sequentially, one after the other. The traced rays from all the shells are then combined to estimate various performance parameters of a multi-shell telescope. In a multi-shell telescope, only rays achieving dual reflections within a single shell, specifically those reflecting first from the primary mirror and then from the secondary mirror of the same shell without interacting with other shells, are considered. Rays that directly reach the detector, those that reach it with a single reflection, or those that reflect more than twice are disregarded as unwanted rays. These unwanted rays can be stopped by incorporating a mechanical baffle/stopper in the telescope design and utilizing mirror shells with non-reflective outer surfaces for X-rays.

In a multi-shell telescope, while all shells may have different focal lengths, they all have a common focus for on-axis rays at the origin. However, the focal plane at which all on-axis rays are focused might not provide the sharpest PSF for off-axis sources as well. To address this, the software allows the detector to be placed at different axial locations ($x_d$). This capability enables users to estimate the variation in PSF size with respect to the detector's axial position for an off-axis source. Furthermore, DarsakX can determine the optimal curvature for a non-planar detector, improving the angular resolution for off-axis sources to address the field curvature associated with the optics.

\subsection{Software Architecture}

DarsakX consists of two main components: geometrical ray tracing and post-processing. The geometrical ray tracing component heavily relies on the telescope's geometrical properties, accounting for factors such as base surface and figure surface in X-ray optics and the X-ray source position ($\theta$ and $\phi=0$). This component manages the bulk of the computational load and remains independent of the X-ray photon's energy and the mirror's reflectivity properties. The geometrical ray tracing part is discussed in Section \ref{sec:01}.

The post-processing segment of DarsakX estimates various telescope performance parameters, including the PSF, Energy Encircled Fraction (EEF), effective area, and the optimal shape of a curved detector to achieve best focus for off-axis sources. This component considers the source's energy and the mirror's reflectivity, utilizing the results obtained from geometrical ray tracing as input parameters. The reflectivity of a surface can be calculated using various tools such as DarpanX \citep{mondal2021darpanx}, IMD \citep{windt1998imd}, GenX \citep{bjorck2007genx} and few more. In DarsakX, we utilized DarpanX to calculate the reflectivity of the surface. The post-processing part is discussed in Section \ref{sec:02}.

\begin{figure*}
\begin{center}
\includegraphics[width=0.7\linewidth]{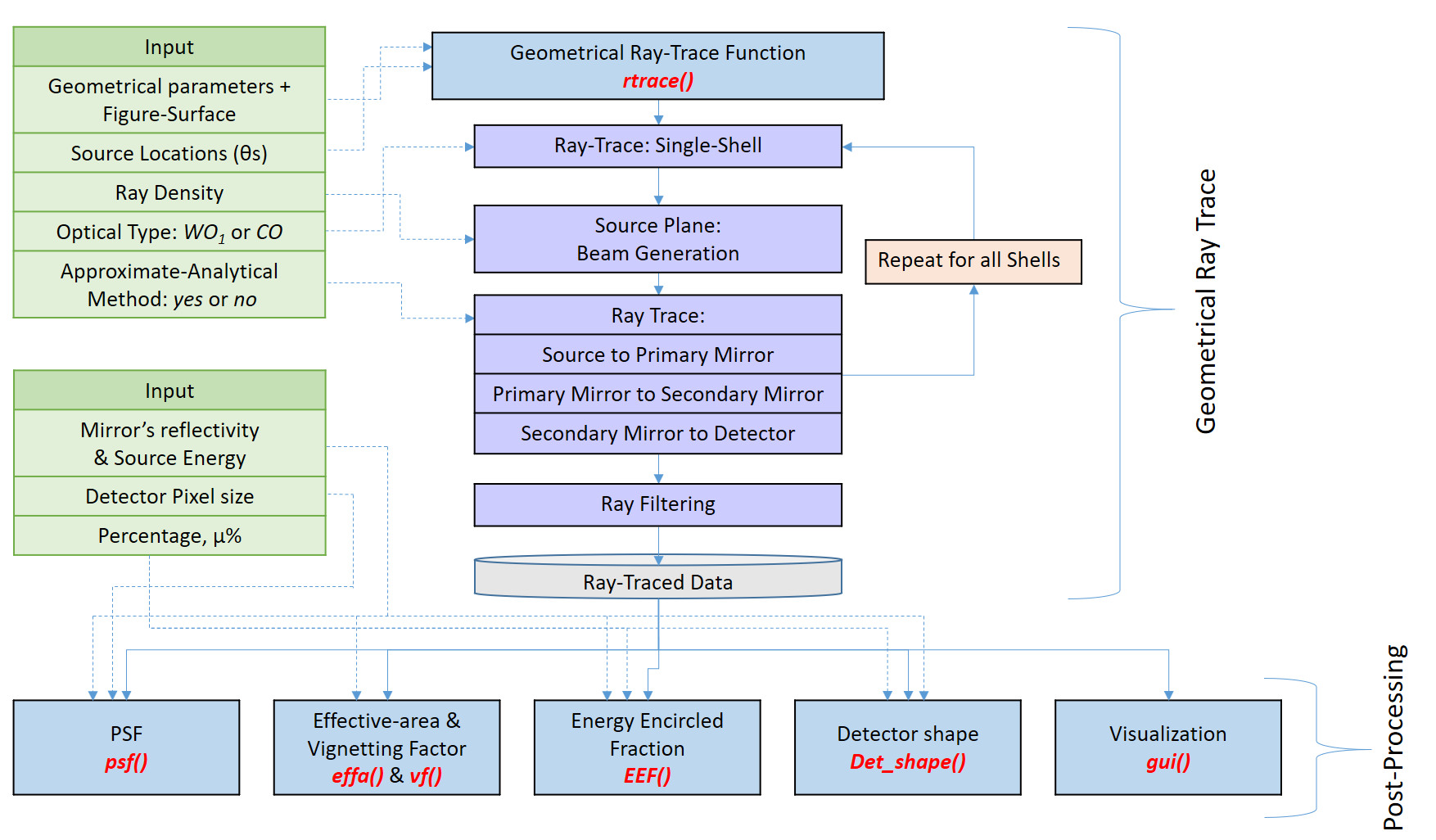}
  \caption{Functional workflow of DarsakX}
  \label{fig:smartdraw}
\end{center}
\end{figure*}

The functional workflow of DarsakX is shown in Figure \ref{fig:smartdraw}, which illustrates both the geometrical and post-processing parts. The software enables simultaneous ray tracing for various source locations by providing different values of $\theta$. DarsakX also offers parallel processing capabilities for the geometrical ray tracing section to distribute the workload for different $\theta$ values efficiently. DarsakX includes a Python class called \textit{rtrace()} responsible for computing the geometrical ray tracing part. This function requires input parameters regarding telescope geometry, which are explained in Figure \ref{fig:smartdraw}. The output of this function provides the geometrical path of each ray. The same Python class also incorporates various methods shown in the post-processing section of Figure \ref{fig:smartdraw}. These methods, such as effective area (\textit{effa()}), vignetting factor (\textit{vf()}), point spread function (\textit{psf()}), energy encircled fraction (\textit{EEF()}), detector shape (\textit{det\_shape()}), and visualization  (\textit{gui()}), utilize the geometrical ray traced data obtained through \textit{rtrace()}. 

DarsakX is available as an open-source Python package at GitHub\footnote{https://github.com/xastprl/darsakx.git}. More details regarding the input and output formats of all these functions can be found in the DarsakX user manual.

\section{Ray-Tracing Algorithm}
\label{sec:01}
\subsection{Geometrical Ray-Tracing Through Single Shell}

\subsubsection{Base Surface as $WO_1$}
\begin{figure*}
\begin{center}
  \includegraphics[trim=1cm 0cm 0cm 0cm, clip,width=0.75\linewidth]{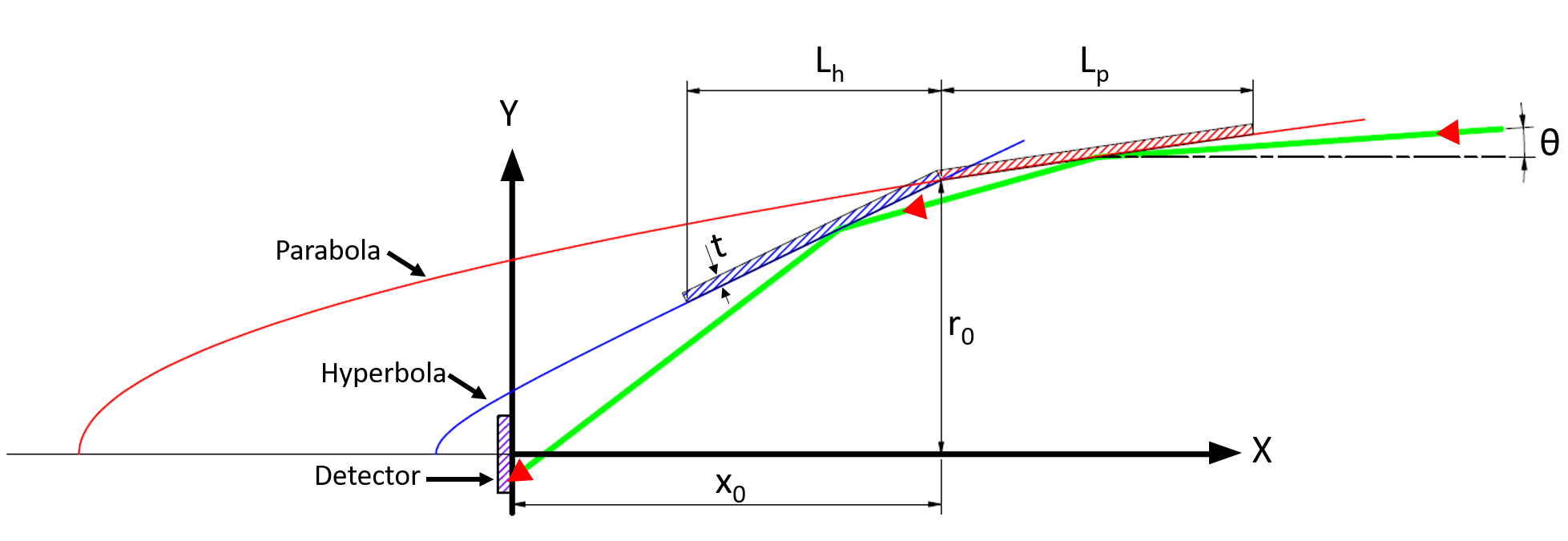}
  \caption{Sectional view of Wolter-1 Optics}
  \label{fig:A1_0}
\end{center}
\end{figure*}

The Wolter-1 optics consist of two axially symmetric mirrors: the paraboloid section serves as the primary mirror for the first reflection, and the hyperboloid section serves as the secondary mirror for the second reflection, as shown in Figure \ref{fig:A1_0}. 

\begin{figure}[h]
\begin{center}
  \includegraphics[width=1\linewidth]{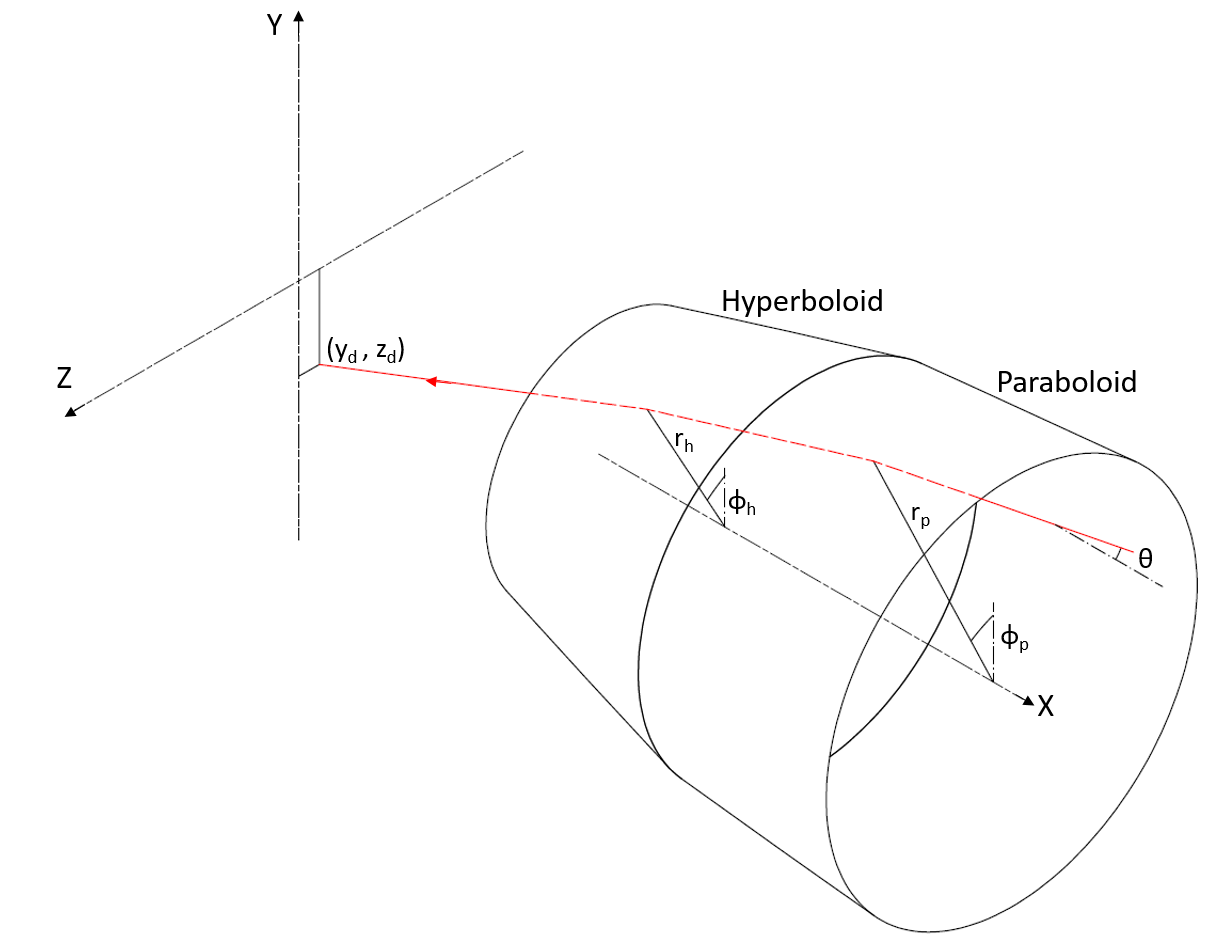}
  \caption{Schematic diagram for off-axis  Wolter-1 configuration}
  \label{fig:A1}
\end{center}
\end{figure}

In DarsakX, a single shell can be considered in a $WO_1$ configuration, and the user can define a $WO_1$ shell using these key parameters: $x_0$, $r_0$, $L_h$, $L_p$, $t$, and $\xi$. The focal length of the shell, denoted as $x_0$, is defined as the distance along the X-axis between the focus of the hyperboloid mirror and the intersection plane formed by both mirrors. The radius of the shell, $r_0$ is determined by the radius of the circle formed at the intersection of both mirrors. The projected length of the paraboloid and hyperboloid surfaces along the X-axis are represented by $L_p$ and $L_h$, respectively. The thickness of the shell is denoted as $t$. Additionally, $\xi$ describes the ratio of two angles, $\alpha_{p0}/\alpha_{h0}$. Here, $\alpha_{p0}$ represents the incident angle that an on-axis ray makes with the primary mirror at a location very close to the intersection of the two mirrors. The $\alpha_{h0}$ denotes the angle that the same ray makes with the secondary mirror at a location very close to the intersection of both mirrors, after it has been reflected from the primary mirror. The surface of $WO_1$ with the figure surface can be defined by the Equation \ref{eq:A1}-\ref{eq:A4}. Where vector  $\vec{R}_{p}$ and $\vec{R}_{h}$ define the paraboloid and hyperboloid surface as shown in Figure \ref{fig:A1}.

\begin{equation}  \label{eq:A1}
\vec{R}_{p}=[x_{p}, r_{p}\cos\phi_{p}, r_{p}\sin\phi_{p}], \hspace{4pt} x_{p} \in [x_{0}, x_{0}+L_{p}] \hspace{2pt} and \hspace{2pt} \phi_{p} \in [0,2\pi]
\end{equation}

\begin{equation}  \label{eq:A2}
\vec{R}_{h}=[x_{h}, r_{h}\cos\phi_{h}, r_{h}\sin\phi_{h}], \hspace{4pt} x_{h} \in [x_{0}-L_{h}, x_{0}] \hspace{2pt} and \hspace{2pt} \phi_{h} \in [0,2\pi]
\end{equation}

\begin{equation}  \label{eq:A3}
r_{p}=\sqrt{p^2+2px_{p}+4e^2pd/(e^2-1)}+G_{p}(x_{p})
\end{equation}

\begin{equation}  \label{eq:A4}
r_{h}=\sqrt{ e^2(x_{h}+d)^2-x^2} + G_{h}(x_{h})
\end{equation} 

The focal point of the optics is defined at $x=0$. The Equation \ref{eq:A3} and \ref{eq:A4} are the equation of the paraboloid and hyperboloid surface as described in \cite{vanspeybroeck1972design} along with the figure surface $G_{p}(x_{p})$ and $G_{h}(x_{h})$,  present in the paraboloid and hyperboloid section in axial direction.  It is assumed that the shell is axially symmetric; therefore, the surface figures ($G_{p}(x_{p})$ and $G_{h}(x_{h})$) only depend on $x$ and are independent of $\phi$. The physical significance of $e$, $p$, and $d$ can be obtained from the study by  \cite{vanspeybroeck1972design}. In DarsakX, the values of $e$, $p$, and $d$ are computed based on the user-provided input parameters: $x_0$, $r_0$, and $\xi$.

In \ref{sec:Appendix}, we outline the procedure for tracing a single ray that originates from the source plane mesh grid and undergoes two reflections within a single shell, ultimately reaching the detector plane.

\subsubsection{Base Surface as $CO$}
\label{sec:03}
In the conical approximation, we consider conical surfaces instead of paraboloid and hyperboloid surfaces. The axial profile of the paraboloid and hyperboloid surfaces, which are represented by Equation \ref{eq:A3} and Equation \ref{eq:A4}, will be replaced as follows:

\begin{equation}  \label{eq:A46}
r_{p}=\left( r_0+\tan\beta_{p0}\left(x-x_0\right) \right) +G_{p}(x_{p})
\end{equation}

\begin{equation}  \label{eq:A47}
r_{h}=\left( r_0+\tan\beta_{h0}\left(x-x_0\right) \right) + G_{h}(x_{h})
\end{equation} 

From here, the paraboloid surface will be represented as the primary mirror, and the hyperboloid surface will be represented as the secondary mirror, specifically within the context of the conical approximation method.

The determination of the slope angles for the primary and secondary mirror, denoted as $\beta_{p0}$ and $\beta_{h0}$ respectively, raises the question of whether they can be considered identical to the slope angles of the parabola and hyperbola sections at their intersection in the $WO_1$ case. The $\beta_{p0}$ and $\beta_{h0}$,  derived in $WO_1$ case, are the slope angles of the paraboloid and hyperboloid surfaces at their intersection (at $x=x_0$). These values are derived from the requirement that when a ray hits the paraboloid section very close to the intersection ($x=x_0$) and finally reaches the detector, it must hit the center of the detector at $\vec{R}=[0,0,0]$.

In contrast to the $WO_1$, the on-axis PSF for the conical approximation is not a delta function with respect to geometrical optics (when diffraction is ignored). The on-axis PSF is determined by the values of $\beta_{p0}$ and $\beta_{h0}$. However, if the same value of $\beta_{p0}$ and $\beta_{h0}$, which are considered for $WO_1$ case, are used for the conical case as well then it will not produce the sharpest PSF. The sharpest on-axis PSF will be produced when a ray, which hits the primary mirror at $x=x_0+l_p/2$, reaches the center of the detector.

To achieve the optimal on-axis PSF, one can find the values of $\beta_{p0}$ and $\beta_{h0}$ analytically by imposing the constraint that a ray hitting the primary mirror at $x=x_0+l_p/2$ should reach the center of the detector. However, the equations required to obtain the analytical solution are non-linear and complex, making numerical solutions necessary.

An alternate method to find an approximate solution for $\beta_{p0}$ and $\beta_{h0}$ is to consider a 2D section of the conical shell at $\phi=0$. Instead of requiring that a ray striking the primary mirror at $x=x_0+l_p/2$ needs to reach the detector at $\vec{R}=[0,0,0]$, we can define that rays striking the primary mirror at $x=x_0$ and at $x=x_0+l_p$ should hit the detector plane at $\vec{R}=[0,-r,0]$ and $\vec{R}=[0,r,0]$, respectively. Here, $2r$ represents the diameter of the PSF, where $r\approx l_p\tan\beta_{p0}\sqrt{{x_0}^2+{r_0}^2}/2x_0$. Hence updated $\beta_{p0}$ and $\beta_{h0}$ can be calculated as follows.

\begin{equation}  \label{eq:A48}
\beta_{p0}=\arctan\left(\frac{r_0+r}{x_0}\right)\frac{\xi}{2(\xi+1)}, \hspace{4pt} and  \hspace{4pt} \beta_{h0}=\frac{(2\xi+1)\beta_{p0}}{\xi}
\end{equation} 
 
The methodology of ray tracing through a shell with the base surface as $CO$ is detailed in the \ref{sec:Appendix_CO}.

\subsubsection{Effect of Figure Surface: An Approximate-Analytical Method} \label{sec:A3}

A ray tracing process involves multiple steps, such as determining the intersection coordinates of the ray with the primary and secondary mirror surfaces and the detector, as well as tracking ray directions after reflection from the primary and secondary mirrors. In instances where a shell with a base surface of either $WO_1$ or $CO$ has a zero figure surface ($G_{p}(x_{p})=0$ and $G_{h}(x_{h})=0$) for both the primary and secondary mirrors, we efficiently solve all relevant equations analytically (\ref{sec:Appendix} and \ref{sec:Appendix_CO}). However, in cases where a mirror shell possesses a non-zero figure surface ($G_{p}(x_{p})\neq0$ and $G_{h}(x_{h})\neq0$), the intersection coordinates of the ray with the primary and secondary mirrors cannot always be analytically solved and hence necessitate a numerical solution. Solving the non-linear equations numerically significantly increases the computational load, thereby leading to longer processing times. Furthermore, when evaluating the telescope's performance across a large number of shells involving multiple sets of $\theta$ and energy with a high density of rays, the computational time may increase substantially.

We have developed an approximate-analytical method (AAM) by which the trajectory of the ray can be solved analytically, even when non-zero figure surfaces ($G_{p}(x_{p})\neq0$ and $G_{h}(x_{h})\neq0$) are present in a mirror shell along with the base surface. In this method, while estimating the intersection of the ray with the mirror surface, the figure surface is considered to be zero ($G_{p}(x_{p})=0$ and $G_{h}(x_{h})=0$), and the intersection of the ray is considered only with the base surface. The effect of the figure surface is considered only when estimating the direction of the ray after reflection. This is because, in the case of X-ray optics, angular resolution is more sensitive to the error associated with the direction of the reflected ray than the error associated with the position from where the ray gets reflected. This method is only applicable when the figure surface is very small (typically around a micron level) compared to the base optics. The methodology for tracing rays through a single shell using the AAM is described in \ref{sec:Appendix_3}. In DarsakX, we have provided the option to trace the ray using both exact and approximate methods.

\subsection{Geometrical Ray-Tracing Through Multiple Shell} \label{sec:mutishell}

A multi-shell telescope consists of multiple mirror shells that are coaxially aligned. Where each shell can be defined as a combination of the base surface (like $WO_1$ or $CO$) and figure surface ($G_{p}(x_{p})$ and $G_{h}(x_{h})$).

In a multi-shell telescope where all shells have the same focal length, a common issue arises. When these shells, having the same focal length, share the same on-axis focal plane but have different radial positions, they may not yield optimal angular resolution for off-axis sources due to their varying plate scales \citep{saha2022optical}. For a telescope with multiple shells, the principal surface where the primary and secondary mirrors intersect for each shell should form a circle. In DarsakX, we have provided an option where multiple shells can be positioned in such a way that they share a common on-axis focal plane but have different focal lengths.

\begin{figure}[h]
\begin{center}
  \includegraphics[scale=0.32]{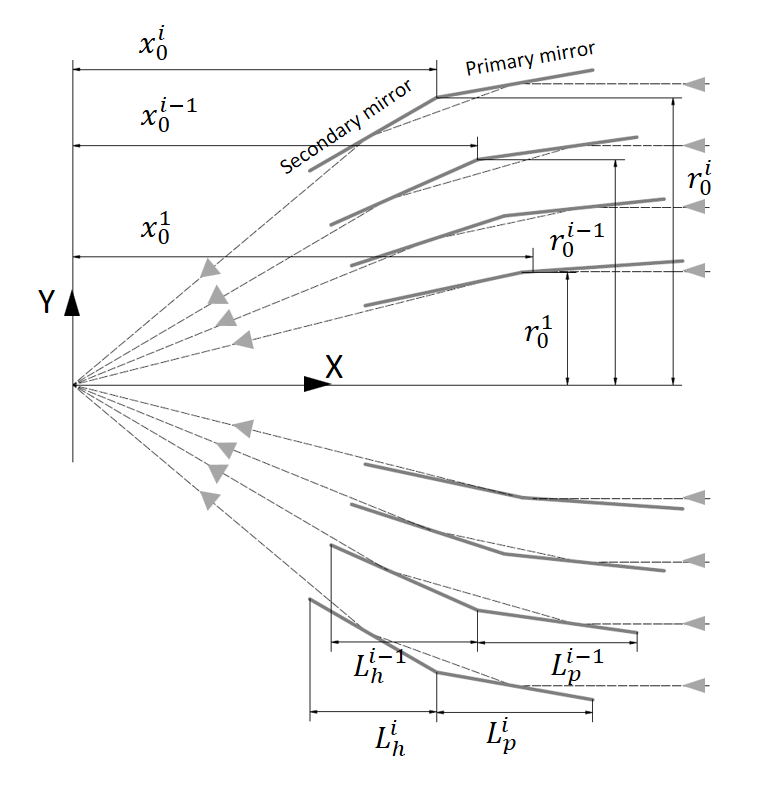}
  \caption{Geometrical parameters of a multi-shell telescope.}
  \label{fig:multishell}
\end{center}
\end{figure}

Figure \ref{fig:multishell} represents a sectional view demonstrating on-axis ray tracing within a multi-shell telescope. Each shell, identified as the $i_{th}$ shell, has various geometric parameters: the type of base surface ($WO_1$ or $CO$), figure surface ($G^i_{p}(x_{p})$ and $G^i_{h}(x_{h})$), inner radius ($r^i_0$), focal length ($x^i_0$), lengths of the primary and secondary mirrors ($L^i_p$, $L^i_h$), shell thickness ($t^i$), and $\xi^i$. Considering the shell thickness
is crucial, as inadequate spacing between shells can impact the effective area.

\begin{figure}
\begin{center}
\includegraphics[width=0.8\linewidth]{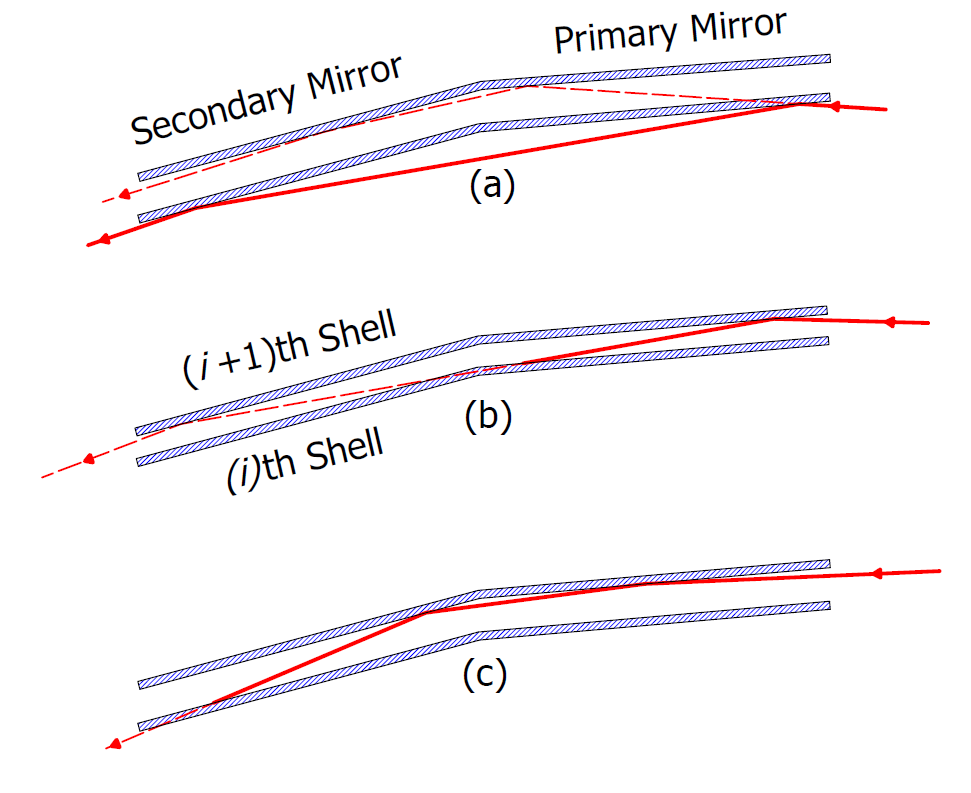}
  \caption{Schematic Diagram of Ray Filtering}
  \label{fig:rayfilter}
\end{center}
\end{figure}

Ray tracing in a multi-shell telescope involves tracing rays from a point source at infinity to the focal plane through each shell individually. Subsequently, the rays from all shells are combined to evaluate various telescope parameters. However, if the gap between two consecutive shells is too small, the $i_{th}$ shell might obstruct some rays intended for the $(i+1)_{th}$ shell, and vice versa. Therefore, the rays have been filtered according to the following cases, which are also illustrated in Figure \ref{fig:rayfilter}.

\begin{enumerate}[label=(\alph*)]
\item Rays that hit the inner surface of the primary mirror of the $i_{th}$ shell should be excluded from the traced rays of the $(i+1)_{th}$ shell.
\item Rays that hit the inner surface of the primary mirror of the $(i+1)_{th}$ shell, and then hit the outer surface of the primary or secondary mirror of the $i_{th}$ shell before reaching the inner surface of the secondary mirror of the $(i+1)_{th}$ shell, should be excluded.
\item Rays that hit the inner surface of the secondary mirror of the $(i+1)_{th}$ shell, and then hit the outer surface of the secondary mirror of the $i_{th}$ shell, should be excluded.
\end{enumerate}

\subsection{Geometrical Ray Trace: Input, Output, and Usage}

During the geometrical ray trace process in DarsakX, the setup of the multi-shell X-ray telescope must be provided as input parameters. These input parameters involve defining the surface of each shell, which can be specified by either $WO_1$ or $CO$ as the base surface, along with the figure surface ($G_{p}(x_{p})$ and $G_{h}(x_{h})$), and the derivative of the figure surface with respect to  $x$ axis ($G'_{p}(x_{p})$ and $G'_{h}(x_{h})$). Additionally, the parameters $r_0$, $x_0$, $L_p$, $L_h$, $t$, and $\xi$ are also necessary to define the base surface for each shell.

Multiple sources can be simultaneously traced by assigning various $\theta$ values. The ray density ($\mathrm{rays/cm^2}$), denoted as $\rho$, also needs to be provided as an input parameter. The axial position of the detector, denoted as $x_d$, is another crucial input. Users have the option to trace rays using the AAM to reduce computational time. All these essential input parameters for geometrical ray tracing are detailed in Table \ref{tab:T1}.

The output of the geometrical ray trace provides specific details for each ray: the incoming ray's direction onto the primary mirror, the coordinates where the ray intersects the primary mirror, the direction of the reflected ray from the primary mirror, the intersection of the ray with the secondary mirror, direction of reflected ray from secondary mirror, and the ray's position on the detector plane. Furthermore, output parameters include the incident angle of the ray on the reflecting surface ($\alpha_p$ with the primary mirror and $\alpha_h$ with the secondary mirror). These incident angles are crucial in the post-processing phase to estimate the reflectivity of the ray upon reflection from the mirror.

It is essential to note that the entire geometrical ray tracing process remains independent of the energy of the incident photon. However, the reflectivity of the ray depends on both the photon's energy and the incident angle. Thus, these incident angles serve as inputs for estimating reflectivity in the post-processing section. The incident angles $\alpha_p$ and $\alpha_h$ for a ray can be calculated using the following equations:

\begin{equation}  \label{eq:A38}
\alpha_p=\arccos(\hat{n}_{\perp p}.\hat{n}_{ip})-90^\circ
\end{equation} 

\begin{equation}  \label{eq:A39}
\alpha_h=\arccos(\hat{n}_{\perp h}.\hat{n}_{rp})-90^\circ
\end{equation} 

Here, $\hat{n}_{\perp p}$, $\hat{n}_{ip}$, $\hat{n}_{\perp h}$ and $\hat{n}_{ih}$ are described in \ref{sec:Appendix}. Both the input and output parameters of the ray trace process are summarized in Table \ref{tab:T1}. 

\begin{table}[h]
\begin{tabular}{p{0.65\linewidth}p{0.24\linewidth}}
\hline 
\textbf{Input Parameters} & \textbf{Symbol}  \\ 
\hline 
Base  surface & $WO_1$ or $CO$  \\ 
 
Radius of each shell & $r_0$ \\ 

Focal length of each shell & $x_0$  \\ 

Length of primary mirror in each shell& $L_p$  \\ 
 
Length of secondary mirror in each shell & $L_h$  \\ 

Thickness of each shell & $t$  \\ 
 
Incident angle ratio for each shell & $\xi$  \\ 
 
Figure surface and its derivative in primary mirror for each shell & $G_{p}(x_{p})$, $G'_{p}(x_{p})$ \\
 
Figure surface and its derivative in secondary mirror for each shell & $G_{h}(x_{h})$, $G'_{h}(x_{h})$  \\
 
Point source locations & $\theta$  \\ 
 
Ray density & $\rho$  \\ 
 
Detector position& $x_d$ \\
 
Approximate-analytical method & \textit{yes} or \textit{no}  \\ 
\hline 
\textbf{Output parameters} & \textbf{Symbol}\\
\hline 
Direction of incident ray on primary mirror & $[{n_{ip}}_{x}, {n_{ip}}_{y}, {n_{ip}}_{z}]$ \\
 
Ray position on primary mirror& $[x_p$, $r_p$ , $\phi_p]$ \\

Direction of reflected ray from primary mirror & $[{n_{rp}}_{x}, {n_{rp}}_{y}, {n_{rp}}_{z}]$\\
 
Ray position on secondary mirror& $[x_h$, $r_h$ , $\phi_h]$ \\
 
Direction of reflected ray from secondary mirror & $[{n_{rh}}_{x}, {n_{rh}}_{y}, {n_{rh}}_{z}]$ \\

Ray position on detector& $y_d$, $z_d$ \\

Ray incident angle on primary mirror& $\alpha_p$ \\

Ray incident angle on secondary mirror& $\alpha_h$ \\
\hline 
\end{tabular} 
\caption{Input: Telescope geometrical parameters, Output: Ray traced results}
\label{tab:T1}
\end{table}

The geometrical ray trace in DarsakX can be executed using the Python class \textit{rtrace()}. Input and output parameters of the function are the same as described in Table \ref{tab:T1}. The figure surface details, $G_{p}(x_{p})$, $G_{h}(x_{h})$, $G'_{p}(x_{p})$, and $G'_{h}(x_{h})$, must be provided as input to \textit{rtrace()} in Python function format for both mirrors in each shell. Using these functions, DarsakX obtains the figure surface and its derivative for any axial position \textit{x}.

\section{Post-Processing and Telescope Performance}
\label{sec:02}
The data obtained from geometrical ray tracing can be used to assess various performance parameters of X-ray telescopes, such as PSF, EEF, effective area, and the curved detector's shape to minimize field curvature. However, to calculate these performance parameters, it's necessary to estimate the weight of each ray. This weight represents the product of the ray's reflectivities when it strikes the primary and secondary mirrors. The reflectivity of each ray depends on both the incident angle and the energy of the incident photon. The reflectivity curve of a mirror, which provides reflectivity as a function of incident angle and photon energy, is determined by the mirror's coating parameters, such as coating material, whether it is a single layer or multiple layers (and if multiple layers, whether it is Constant period or Depth-graded), coating thickness, thickness uniformity, surface roughness, etc. In DarsakX, DarpanX \citep{mondal2021darpanx} is utilized to calculate the surface reflectivity.

Let's consider that when a monochromatic ray strikes the primary mirror surface with an incident angle of $\alpha_{p}$, it has a reflectivity $R_p(\alpha_{p})$. Similarly, when the same ray strikes the secondary mirror surface with an incident angle $\alpha_{h}$, it has a reflectivity $R_h(\alpha_{h})$. After two reflections, when a ray arrives at the detector, we can define its weight as $w$. It is important to note that different shells may have different coating recipes. Therefore, $R_p$ and $R_h$ can be different for different shells.

\begin{equation}  \label{eq:A40}
w=R_p(\alpha_p)R_h(\alpha_h)
\end{equation} 
 
\subsection{PSF}
If we assume the pixel size of the flat detector to be $l_{pix}$, and we consider a pixel centered at $[x_d, y_{pix}, z_{pix}]$, then the total energy received by this pixel will be given by:

\[E_{pix} = E_{0} \sum_{i} w^{i}\]

\noindent where $E_{0}$ represents the energy associated with each individual ray. The position of the $i_{th}$ ray ($y^{i}_d, z^{i}_d$) on the detector plane satisfies the conditions: 
\[y_{pix} - \frac{l_{pix}}{2} \leq y^{i}_d \leq y_{pix} + \frac{l_{pix}}{2}\]
\[z_{pix} - \frac{l_{pix}}{2} \leq z^{i}_d \leq z_{pix} + \frac{l_{pix}}{2}\]

By calculating the sum of the weights ($w^i$) of all rays that fall within the bounds of the pixel and multiplying it with $E_{0}$, we can determine the total energy ($E_{pix}$) received by that specific pixel. This process is used to generate the image file of the PSF distribution. Each pixel in the image file corresponds to a specific location on the detector plane, and the energy received in that pixel is a result of the combined contributions of all the rays that intersected the corresponding region on the detector.

\subsection{Energy Encircled Fraction}
Energy Encircled Fraction, $EEF_{\mu}=r$, defines that $\mu\%$ of the total energy falls within a radius equal to $r$.

\begin{equation} \label{eq:A41}
\frac{\sum_{r_{i}\leq r} w^{i}}{\sum w^{i}}=\frac{\mu}{100}
\end{equation}

Where $i$ corresponds to the $i_{th}$ ray, and $r^i$ can be calculated as follows:

\begin{equation} \label{eq:A42}
\left( r^{i}\right)^2=\left(z^i_{d}-\overline{z_{d}}\right)^2+ \left(y^i_{d}-\overline{y_{d}}\right)^2
\end{equation}

\begin{equation} \label{eq:A43}
\overline{y_{d}}=\frac{\sum y^i_d w^i}{\sum w^i}, \hspace{4pt} \text{and} \hspace{4pt} \overline{z_{d}}=\frac{\sum z^i_d w^i}{\sum w^i}
\end{equation}

These equations allow us to calculate the energy-encircled fraction, which represents the fraction of energy within a specific radius $r$ on the detector plane. The variables $w^i$, $y^i_d$, and $z^i_d$ correspond to the weight and position of the $i_{th}$ ray on the detector plane, while $\mu$ represents the desired percentage of total energy. The expressions in Equation \ref{eq:A42} determine the radius $r^i$ for each ray, based on its position relative to the centroid coordinates ($\overline{y_{d}}$ and $\overline{z_{d}}$) of all rays contributing to the total energy. 

In X-ray astronomy, angular resolution is generally evaluated using the Half Power Diameter ($HPD$), defined as $HPD=2EEF_{50\%}$.

\subsection{Effective Area and Vignetting Factor}

The effective area $A_{eff}$ can be calculated by summing the effective areas of each ray received at the detector plane.

\begin{equation} \label{eq:A44}
A_{eff}=(l_g)^2 \sum w^i
\end{equation}
Here, $l_{g}$ represents the grid spacing, which can be calculated as $l_{g}=1/\sqrt{\rho}$. The vignetting factor for a source placed at an angle $\theta$ can be calculated as:

\begin{equation} \label{eq:A45}
VF(\theta)=\frac{A_{eff}(\theta)}{A_{eff}(\theta=0)}
\end{equation}

\subsection{Post-Processing Methods}
A python class \textit{rtrace()} in DarsakX used for geometrical ray trace has various methods such as \textit{effa()}, \textit{vf()}, \textit{psf()}, \textit{EEF()}, \textit{det\_shape()}, and \textit{gui()} to evaluate various performance parameters such as effective area, vignetting factor, point spread function, energy encircled fraction, detector shape, and 3D visualization, respectively in post-processing section. 

Each method requires specific input parameters. For instance, \textit{effa()} and \textit{vf()} necessitate the reflectivity property for both mirrors in each shell. By default, effective area is calculated for all values of $\theta$. \textit{EEF()} also demands mirror reflectivity but additionally requires the percentage value of the total energy for the energy-encircled fraction to be calculated, evaluating for all $\theta$ values. Similarly, \textit{det\_shape()} relies on reflectivity property and energy percentage. For \textit{psf()}, the value of $\theta$ and detector pixel size are required in addition to the reflectivity property of the mirrors. Lastly, \textit{gui()} only needs the value of $\theta$ for which rays are to be visualized along with the number of rays.

The reflectivity curve, which defines the mirror's reflectivity and is dependent on both the incident angle and photon energy, serves as a crucial input parameter in DarsakX's post-processing methods. The post-processing methods in DarsakX can evaluate the telescope's performance for one energy level at a time. To assess performance for multiple energies, these post-processing methods can be executed multiple times in a loop, allowing the evaluation of performance across various energy levels.

It's important to note that, for assessing performance across multiple energy levels, only the post-processing method needs to be executed multiple times. The geometrical ray tracing class (\textit{rtrace()}) remains independent of the energy of the photon and does not require repetition. Consequently, the reflectivity curve can be supplied as a function of the incident angle for a fixed energy value. In DarsakX, this reflectivity curve must be provided as discrete data points for each mirror in every shell. The software utilizes these data points to interpolate and establish a function that correlates incident angle to reflectivity.

\section{Validation and Applications}
\label{sec:04}
\subsection{Validation With Chandra}
We utilized DarsakX to simulate a telescope with an ideal configuration of Chandra, without any geometrical errors. Given the high geometrical accuracy of the Chandra telescope, there is not a significant deviation between its actual results and those simulated with the ideal configuration. Thus, comparison of results with DarsakX with Chandra Ray Tracer (ChaRT) \citep{carter2003chart} can be used to validate it.

\begin{figure*}[h]
  \centering
  \begin{subfigure}{0.47\textwidth}
    \includegraphics[trim=180 6 100 50, clip, width=\linewidth]{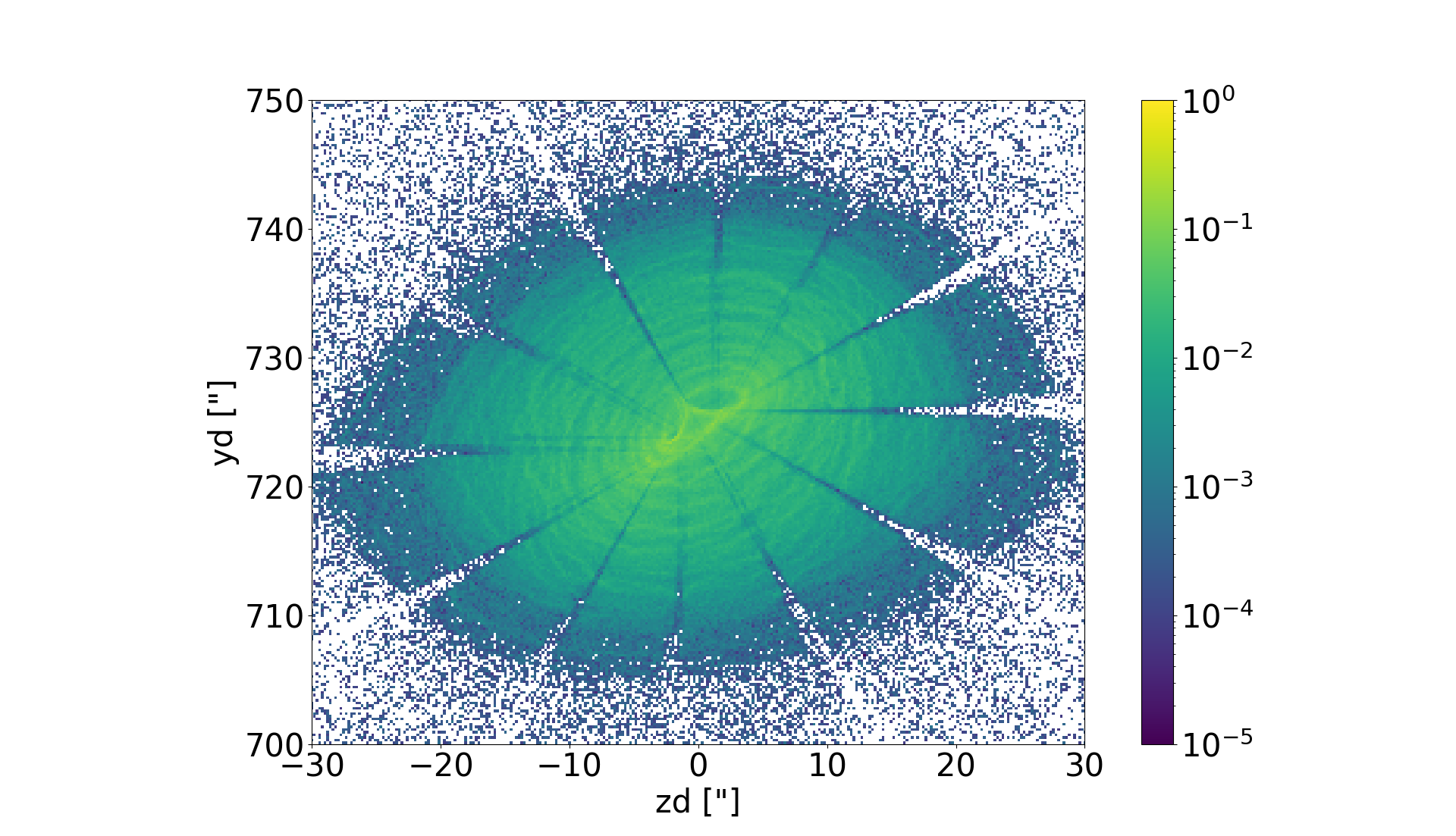}
    \caption{Chandra Ray Tracer}
    \label{fig:A2_1}
  \end{subfigure}
  %\hfill
  %\hspace{0cm}
  \begin{subfigure}{0.47\textwidth}
    \includegraphics[trim=180 6 100 50,clip,width=\linewidth]{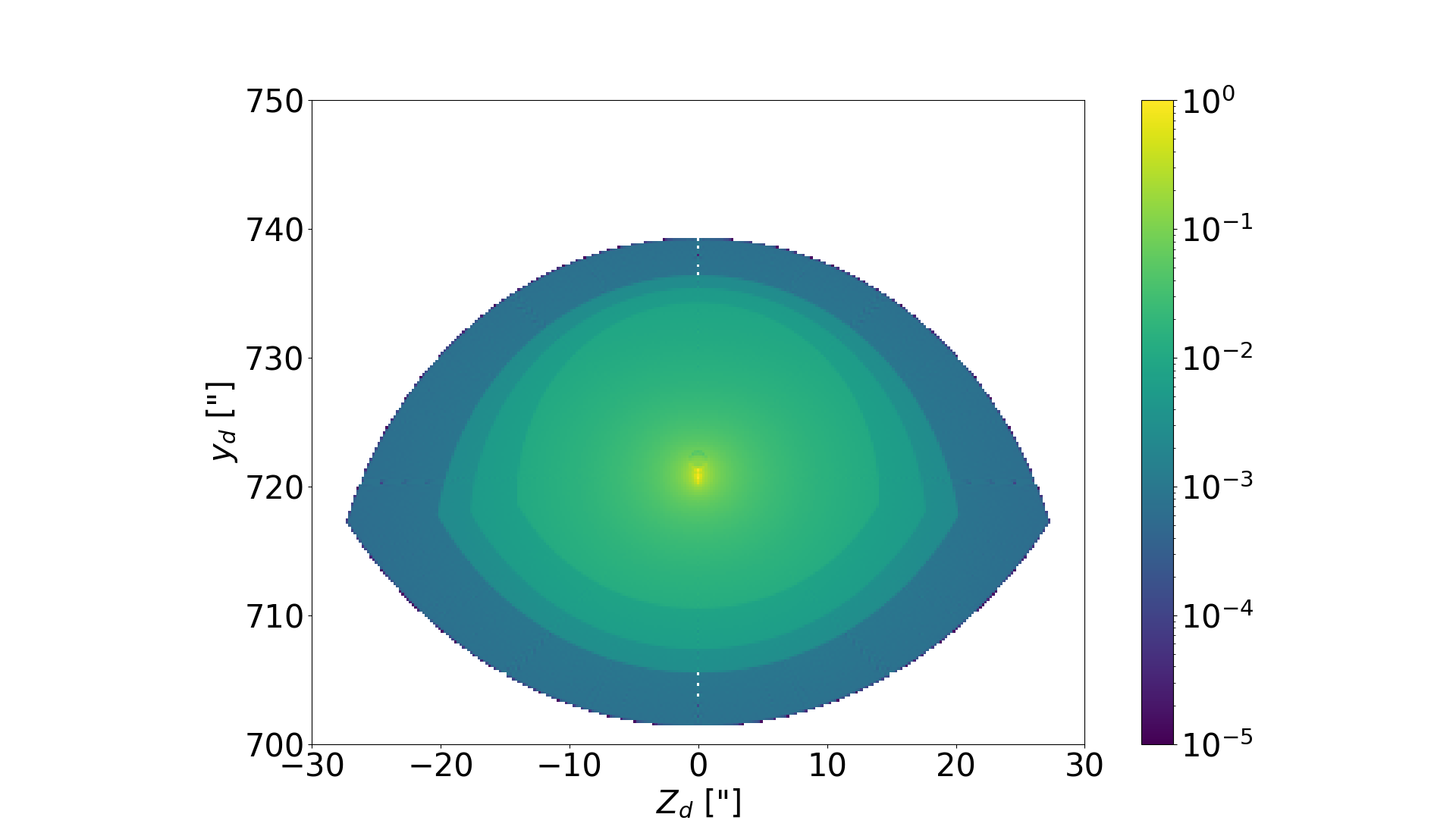}
    \caption{DarsakX}
    \label{fig:A2_2}
  \end{subfigure}
  \caption{(a) PSF simulated by ChaRT, (b)PSF simulated by DarsakX. Source position: $\phi=0^\circ$, $\theta=-0.2^\circ$, energy=4.5 keV and pixel size=10 $\mathrm{\muup m}$.} 
  \label{fig:A2}
\end{figure*}

\begin{figure*}[!h]
\centering
\begin{subfigure}{0.48\textwidth}
  \includegraphics[trim=0cm 0cm 1cm 0cm, clip,width=1\linewidth]{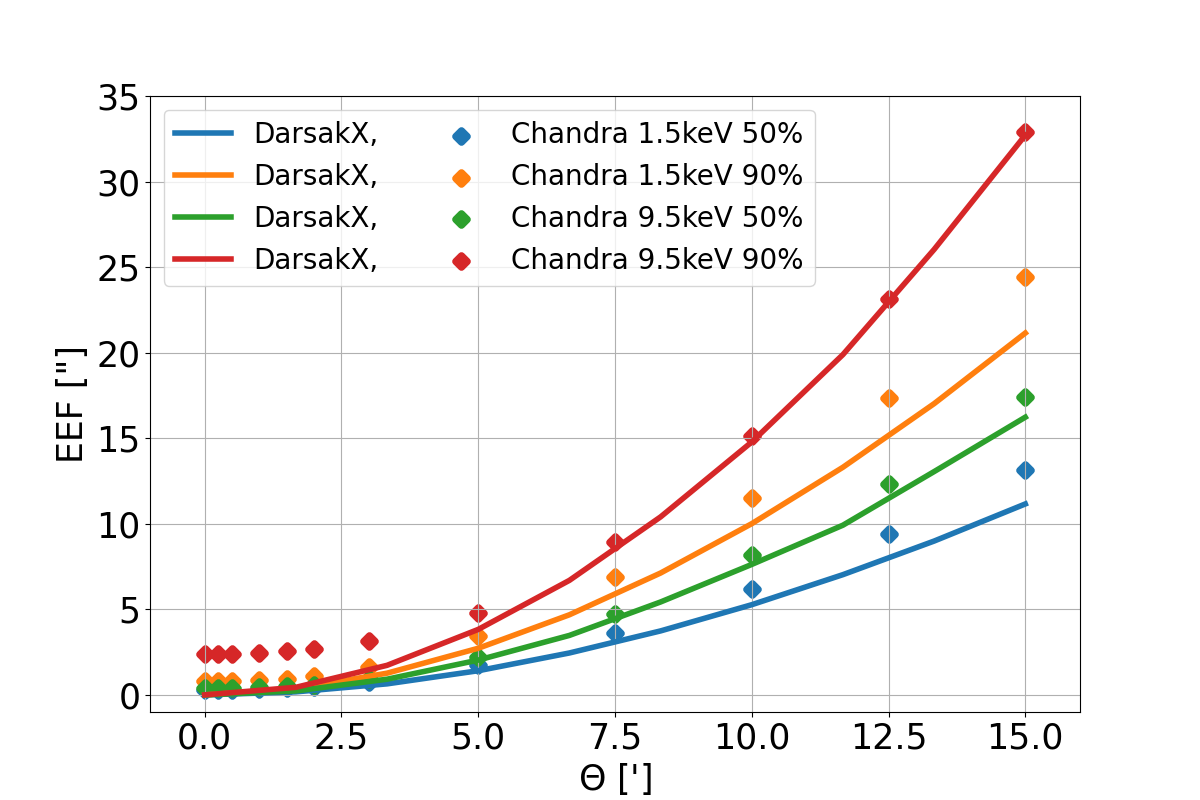}
  \caption{}
  \label{fig:A3A}
\end{subfigure}
\begin{subfigure}{0.48\textwidth}
  \includegraphics[trim=0cm 0cm 1cm 0cm, clip,width=1\linewidth]{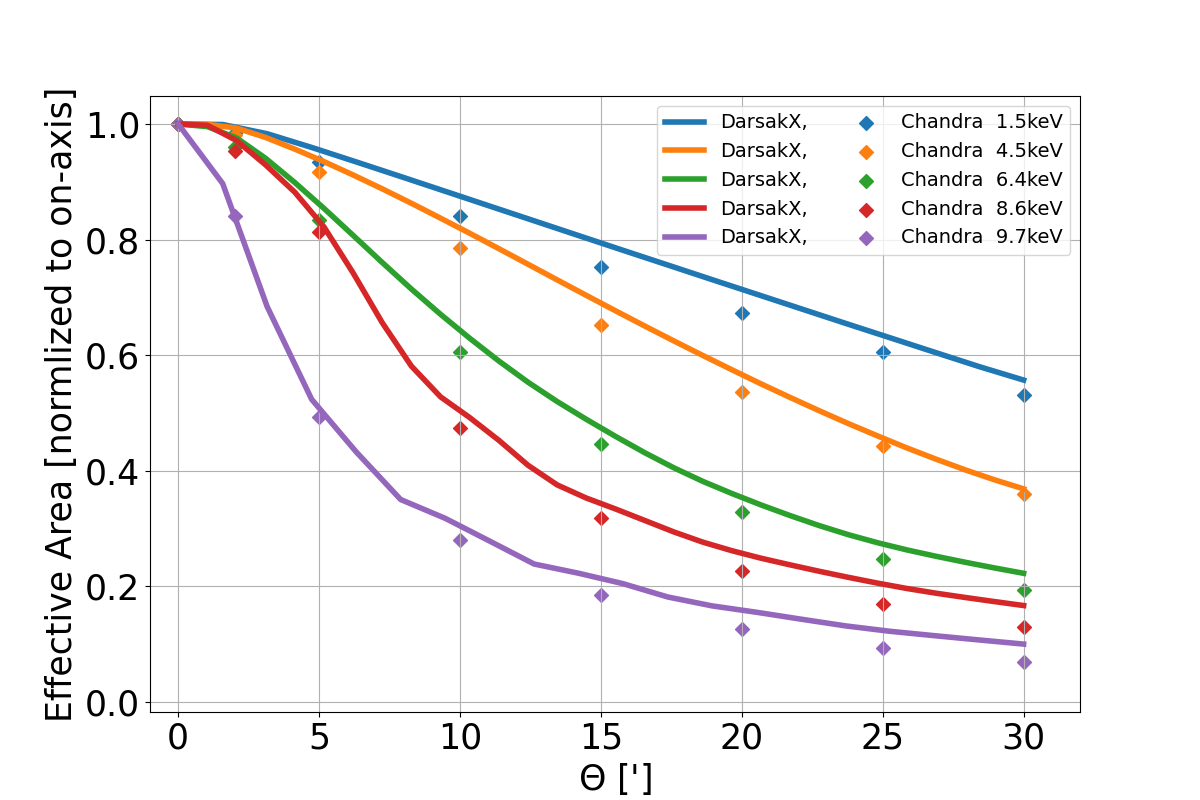}
  \caption{}
  \label{fig:A4A}
\end{subfigure}
  \caption{(a) EEF variation with $\theta$ produced by Chandra-CALDB and DarsakX, (b) Effective area variation with $\theta$ produced by Chandra-CALDB and DarsakX} 
  \label{fig:Chandra_EEF_AREA}
\end{figure*}

Figure \ref{fig:A2} shows the PSF of a source located at $\phi=0$ and $\theta=-0.2^\circ$ with an energy of 4.5 $\mathrm{keV}$, produced by the Chandra telescope. Figure \ref{fig:A2_1} shows the PSF obtained by ChaRT, whereas Figure \ref{fig:A2_2} shows the PSF produced with DarsakX. The overall shape, size, and detector position of the PSF produced by both methods are nearly identical. For computations with DarsakX, we have used the geometrical parameters of Chandra optics from the Chandra Observatory Specifications\footnote{https://chandra.harvard.edu/about/specs.html}. We have not considered the figure error, surface roughness, alignment error, and shadow due to mirror mounting struts present in the geometry. Consequently, the $EEF_{50\%}$ and $EEF_{90\%}$ of the PSF generated by DarsakX are 7.7 arcsec and 14.8 arcsec, respectively. Meanwhile, the $EEF_{50\%}$ and $EEF_{90\%}$ values for the PSF produced by ChaRT are 9.3 arcsec and 17.0 arcsec. This shows that the PSF produced by DarsakX is sharper than that produced by ChaRT, due to DarsakX's utilization for an idealized configuration.

We also compare the EEF and effective area variation with off-axis angles as obtained from DarsakX with the values from Chandra-CALDB 4.10.4 \citep{fruscione2006ciao}, as shown in Figure \ref{fig:Chandra_EEF_AREA}. Since the $WO_1$ produces stigmatic imaging for on-axis sources, in Figure \ref{fig:A3A}, the $EEF$ produced by DarsakX is zero at $\theta=0$. However, the on-axis $EEF_{50\%}$ obtained from Chandra-CALDB is 0.28 and 0.38 arcsec at the energies of 1.5 keV and 9.5 keV, respectively, due to the presence of figure error in the Chandra optics. Chandra's angular resolution at field angles is also affected by figure error but is dominated by optical aberrations present in $WO_1$ itself. Hence, at 15 arcmin, the $EEF_{50\%}$ and $EEF_{90\%}$ produced by DarsakX are 11.2 arcsec and 21.1 arcsec, respectively, at 1.5 keV. Meanwhile, the $EEF_{50\%}$ and $EEF_{90\%}$ obtained from Chandra-CALDB are 13.1 arcsec and 24.4 arcsec, respectively, at 1.5 keV, which are around 14\% higher compared to DarsakX's estimation, attributed to the figure error present in Chandra. 

The vignetting factor obtained from DarsakX and Chandra-CALDB, as shown in Figure \ref{fig:A4A}, closely matches for on-axis sources but shows some discrepancies at higher field angles. Since it is known that figure error primarily affects angular resolution rather than the effective area, these discrepancies likely arise from scattering due to surface roughness on the mirror, shadowing effects caused by various assembly structures, and deviations in the mirror's reflectivity properties from their ideal performance. The vignetting factor estimated by DarsakX at 15 arcminutes is 0.79 and 0.21 at 1.5 keV and 9.7 keV, respectively. Meanwhile, the vignetting factor obtained through Chandra-CALDB is 0.75 and 0.19 at 1.5 keV and 9.7 keV, respectively, which is approximately 5\% and 10\% lower than that estimated by DarsakX.

\subsection{Functionality Examples}
\label{sec:05}
To demonstrate the general capability of DarsakX to simulate the imaging performance of a multi-shell X-ray telescope, we have considered a telescope with 58 shells configured ideally for XMM-Newton, without any geometrical errors. The design parameters for this telescope are listed in Table 1 of \citet{aschenbach2000imaging}.

\begin{figure*}[h]
  \centering
  \begin{subfigure}{0.52\textwidth}
  \includegraphics[trim=3cm 0cm 0cm 0cm, clip, width=1\linewidth]{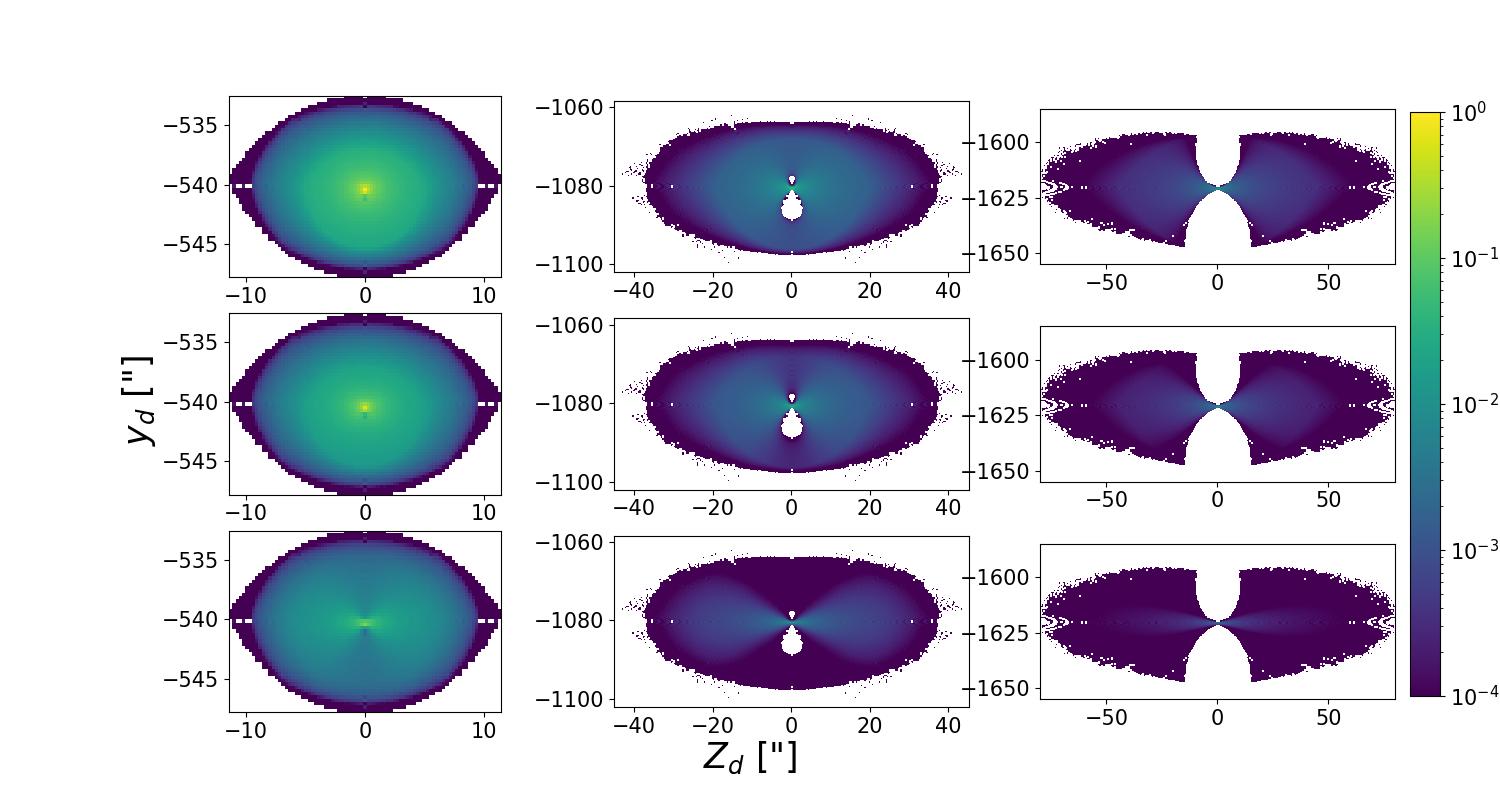}
  \caption{}
  \label{fig:A6}
  \end{subfigure}
  \begin{subfigure}{0.43\textwidth}
      \includegraphics[trim=-1cm 0cm 2cm 0cm, clip, width=1\linewidth]{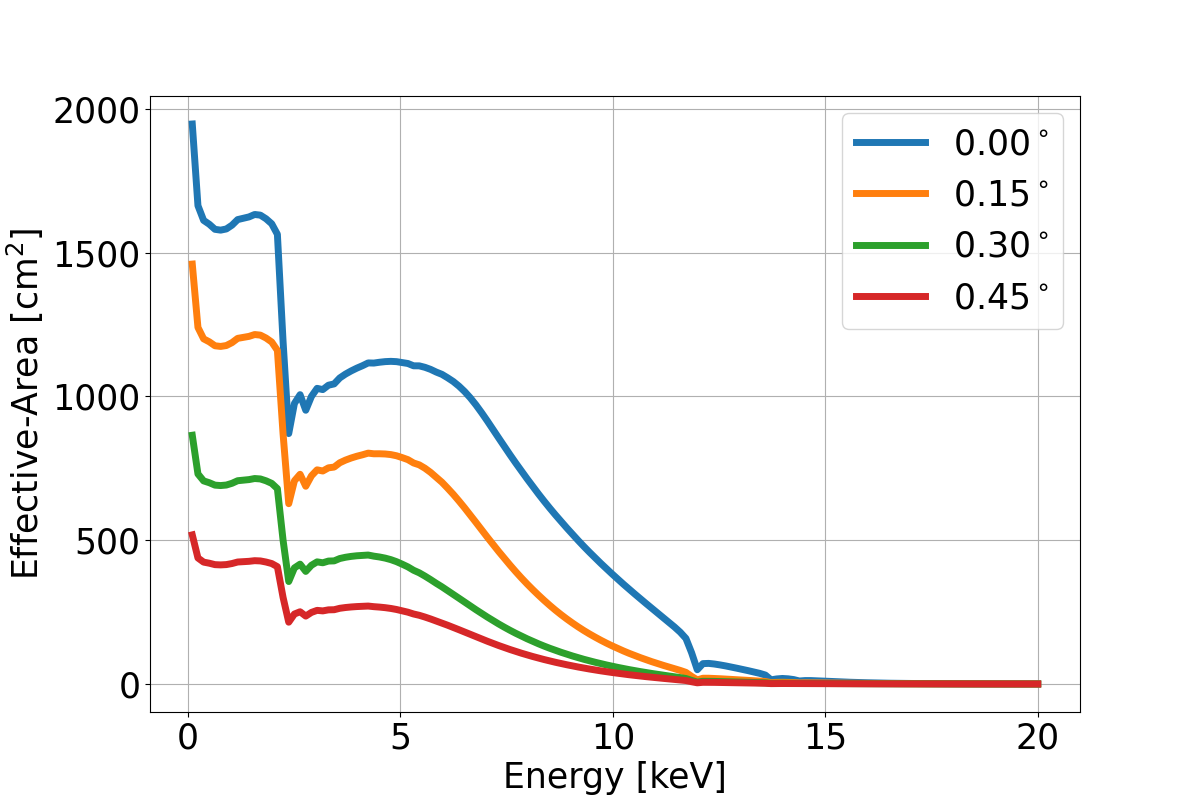}
  \caption{}
  \label{fig:A7}
  \end{subfigure}
  \caption{(a) PSF estimated by DarsakX. The left-to-right columns represent the PSF with variations of $\theta$ at $0.15^\circ$, $0.30^\circ$, and $0.45^\circ$. The top-to-bottom rows represent the PSF with variations of energy at 1 keV, 5 keV, and 8 keV, (b) Effective area variation with energy for few $\theta$, produced by DarsakX.} 
  \label{fig:XMM_PSF}
\end{figure*}

Figure \ref{fig:A6} shows the PSF of the telescope with different sets of energy and off-axis angle $\theta$. The detector pixel size is considered to be 10 $\mathrm{\muup m}$, and the ray density is set to 7500 $\mathrm{rays/cm^{2}}$. Notably, the size and shape of the PSF remain almost independent of energy, while the size increases with higher values of $\theta$. Furthermore, as $\theta$ increases, the overall power becomes more distributed. However, The rise in X-ray energy causes a decrease in the intensity because the mirror's reflectivity decreases with increasing energy. The visible gaps within the PSF, small gaps for $\theta=0.15^\circ$ and relatively larger gaps for $\theta=0.45^\circ$ are a result of shadowing effects caused by the closely packed mirror shells. Figure \ref{fig:A7} displays the effective area variation of the same telescope, produced through DarsakX simulation, as a function of energy for various values of $\theta$.

\begin{figure}[!htb]
\begin{center}
  \includegraphics[trim=1cm 0cm 1cm 0cm, clip,width=0.95\linewidth]{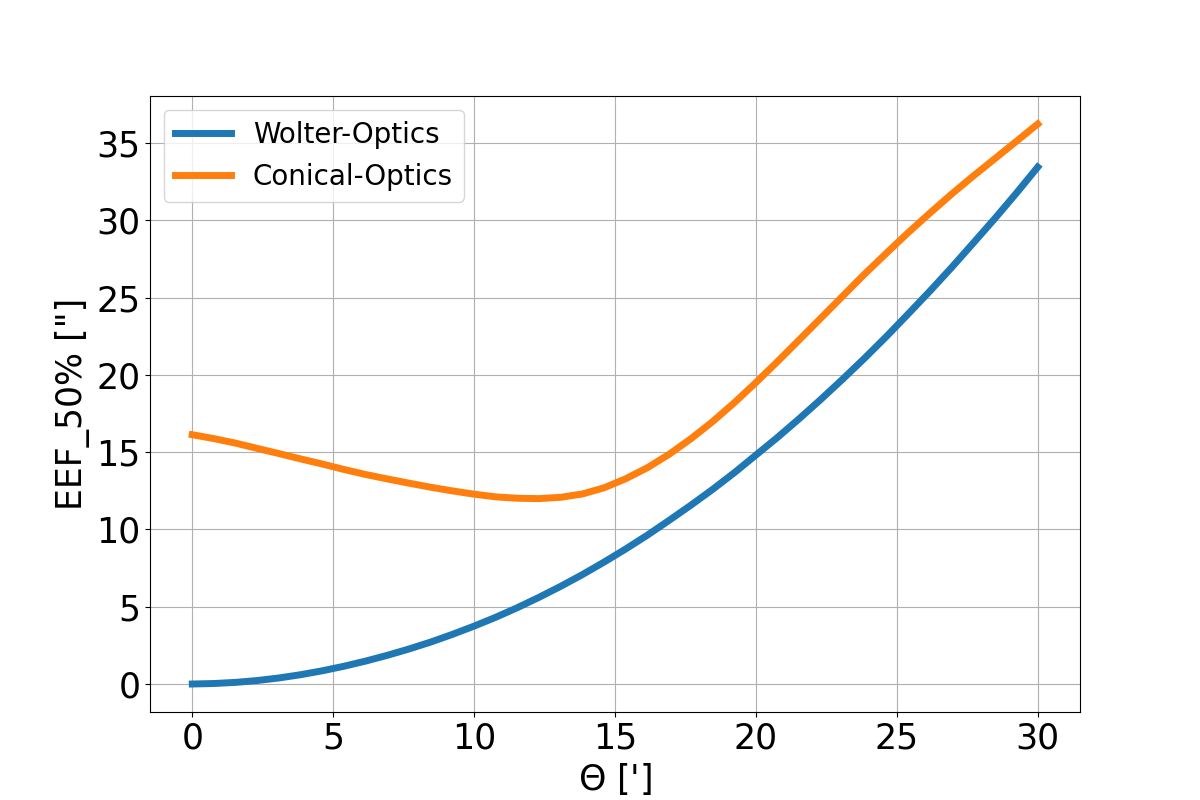}
  \caption{EEF$\_50\%$ variation with off-axis angle $\theta$. Energy of the source is 5 keV, $\theta=0.2^\circ$, detector pixel size 10 $\mathrm{\muup m}$ and ray density 20000 $\mathrm{rays/cm^2}$.}
  \label{fig:A12}
\end{center}
\end{figure}

Another feature of DarsakX is its ability to evaluate the performance of a telescope in both the $WO_1$ and $CO$ configurations. Figure \ref{fig:A12} showcases the variation of $EEF_{50\%}$ with the off-axis angle $\theta$ for both the $WO_1$ and $CO$ configurations. DarsakX can also generate the PSF and other post-processing parameters for the $CO$ configuration.

\subsection{Telescope Performance: Impact of Figure Errors}
\label{sec:approximate-analytica}

\begin{figure}[h]
\begin{center}
  \includegraphics[trim=0cm 0cm 0cm 0cm, clip, width=0.9\linewidth]{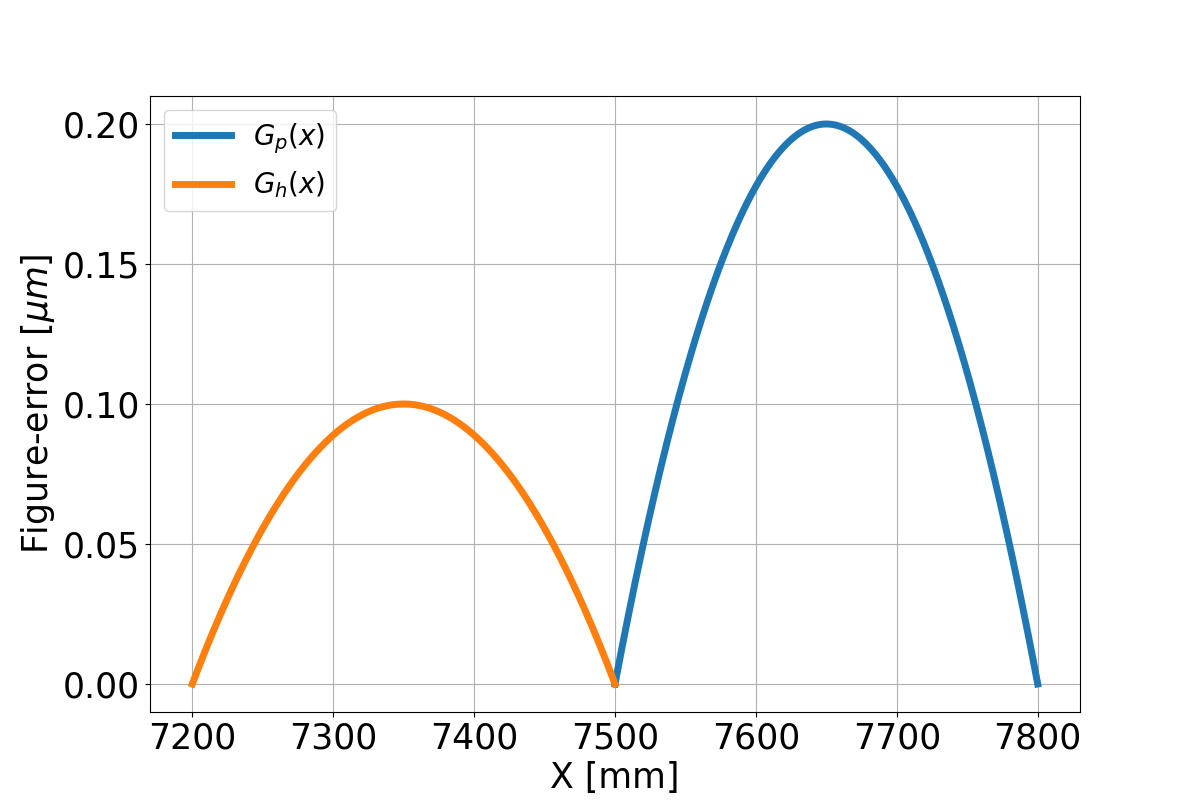}
  \caption{Figure-error profiles, $G_{p}(x_{p})$ and $G_{h}(x_{h})$, present in paraboloid and hyperboloid section}
  \label{fig:A8}
\end{center}
\end{figure}

To demonstrate DarsakX's capability of incorporating figure errors, we consider an ideal telescope configuration introduced in the previous section, with the figure errors included as shown in Figure \ref{fig:A8}. Figure error is considered to be the same in all the shells. These figure errors take the form of parabolic shapes in both the primary and secondary sections. The maximum values of the error are 0.2 $\mathrm{\muup m}$ for the paraboloid section and 0.1 $\mathrm{\muup m}$ for the hyperboloid section.

\begin{figure}[h]
\begin{center}
  \includegraphics[width=1\linewidth]{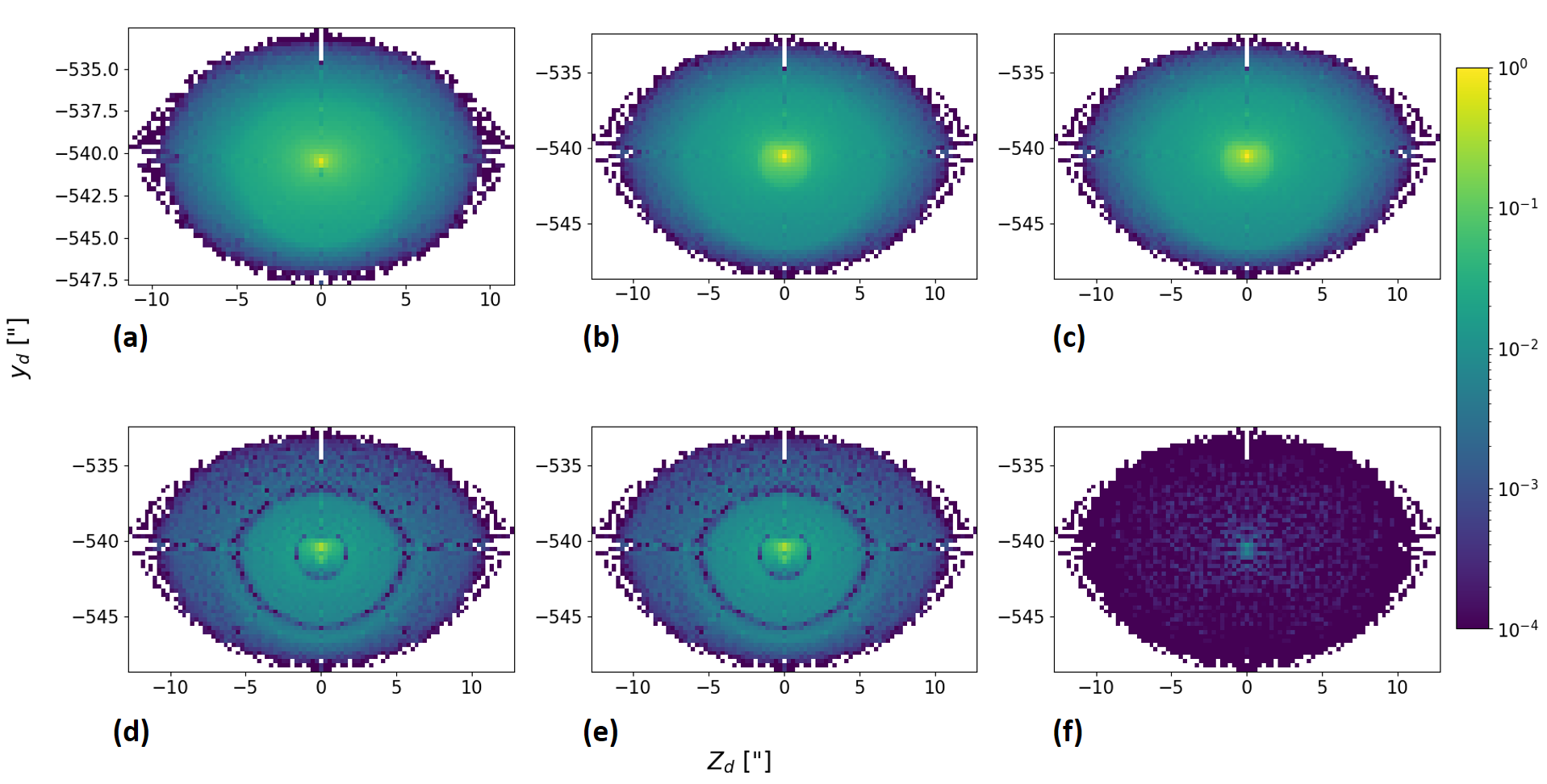}
  \caption{PSF produced by DarsakX for energy 5keV, 
  $\theta=0.15^\circ$ and detector pixel size 10 $\muup \mathrm{m}$. (a) PSF produced for without figure error. (b) PSF produced with the figure error. (c) PSF produced with figure error and by AAM. (d) Difference between \textit{(a)} and \textit{(b)}. (e) Difference between \textit{(a)} and \textit{(c)}. (f) Difference between \textit{(b)} and \textit{(c)}.}
  \label{fig:A9}
\end{center}
\end{figure}

Figure \ref{fig:A9} displays the PSF obtained for three scenarios. In Figure \ref{fig:A9}a, the PSF is generated for the telescope with a perfect $WO_1$ mirror profile, without any figure error. Figure \ref{fig:A9}b illustrates the PSF produced with figure errors (shown in Figure \ref{fig:A8}), simulated using the exact method. Figure \ref{fig:A9}c also shows the PSF produced with figure errors, but simulated using the AAM.

Additionally, Figure \ref{fig:A9}d displays the difference between the PSF produced with figure error, simulated by the exact method, and the PSF produced without figure error (the difference between Figure \ref{fig:A9}b and \ref{fig:A9}a). Figure \ref{fig:A9}e represents the difference between the PSF produced with figure error, simulated by the AAM, and the PSF produced without figure error (the difference between Figure \ref{fig:A9}c and \ref{fig:A9}a). Finally, Figure \ref{fig:A9}f displays the difference between the PSF produced with figure error, simulated by the exact and approximate methods (the difference between Figure \ref{fig:A9}b and \ref{fig:A9}c). This analysis highlights that the difference between the PSFs produced by the exact method and AAM is negligible.

\begin{figure}[h]
\begin{center}
  \includegraphics[trim=0cm 0cm 0cm 0cm, clip, width=1\linewidth]{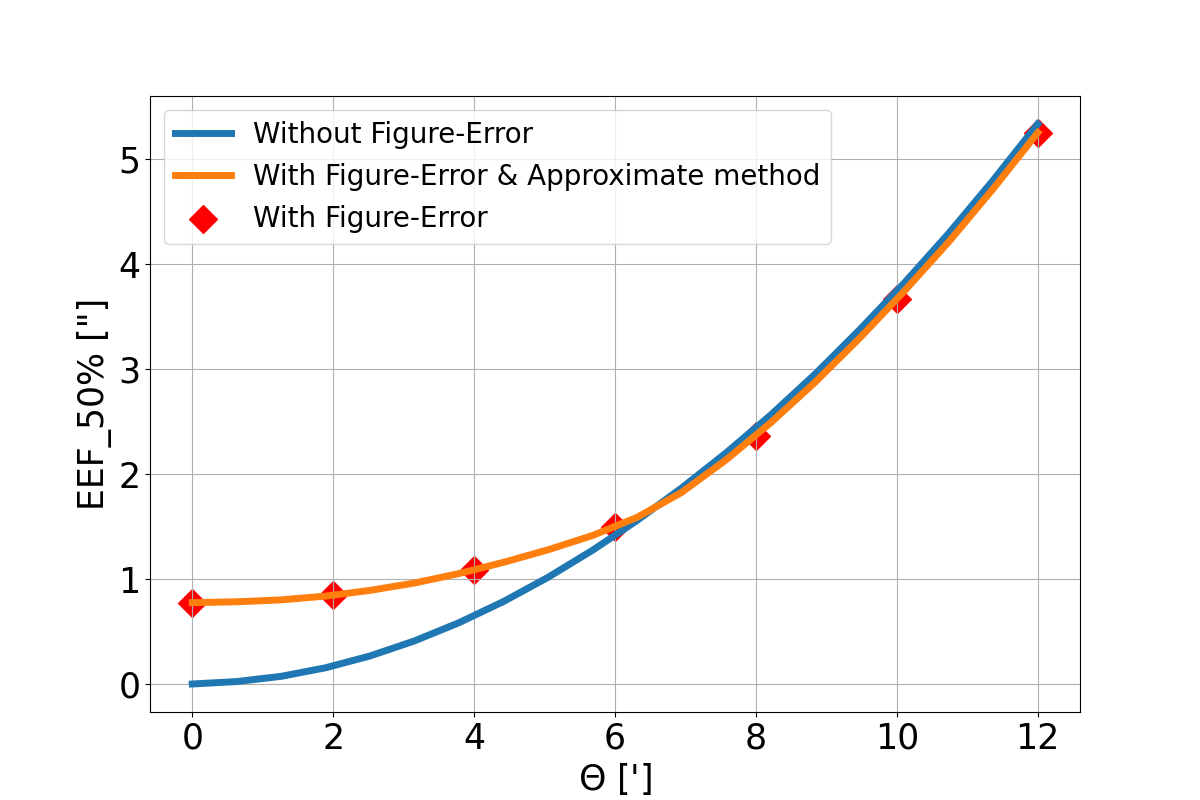}
  \caption{EEF$\_50\%$ variation with off-axis angle $\theta$.}
  \label{fig:A10}
\end{center}
\end{figure}

The variation of $EEF_{50\%}$ for all three cases is presented in Figure \ref{fig:A10}. It is evident that the presence of figure errors predominantly affects on-axis performance, and the $EEF_{50\%}$ values obtained for exact and AAM are nearly identical.

Notably, the computational time required by the exact method is around 180 times higher than that of the AAM. However, this ratio slightly varies with the ray density ($\rho$) and field angle ($\theta$), as shown in Table \ref{tab:runtime}, where the computational times taken by each method and their ratio are presented.

For instance, when the ray density is 5000 $\mathrm{rays/cm^2}$ and $\theta=0.2^\circ$, the computational time taken by the exact method is 10.6 hours, which is 197.2 times higher than the computational time taken by the AAM, i.e., 3.2 minutes. These computation times will further increase with an increase in the number of shells in the telescope, or an increase in the shell diameter or length. Additionally, for the exact method, computational time will further increase if the figure surface is defined by more complex non-linear equations. Therefore, utilizing the AAM is more efficient.

\begin{table}[h]
    \centering
    \begin{tabularx}{\linewidth}{p{1.9cm}XXXX}
    \hline
    \multirow{2}{1.9cm}{\textbf{Ray Density} \textbf{$\rho(\mathrm{rays/cm^2})$}}&\multirow{2}{1.3cm}{\textbf{Method $\&$ Ratio}}&\multicolumn{3}{c} {\textbf{Run-Time}}\\
        &&$\theta=$0$^\circ$&$\theta=$0.1$^\circ$&$\theta=$0.2$^\circ$\\
        \hline
         \multirow{3}{*}{50}&Exact&188.1s&381.7s&389.0s\\
        
        &AAM&1.7s&2.6s&2.6s\\
        
        &Ratio&112.6&144.8&147.6\\
        \hline
                 \multirow{3}{*}{100}&Exact&377.8s&769.5s&795.6s\\
       
        &AAM&2.4s&4.48s&4.3s\\
       
        &Ratio&157.5&173.9&183.9\\
        \hline
                 \multirow{3}{*}{1000}&Exact&3777.1s&7558.7s&7701.9s\\
        
        &AAM&16.6s&35.6s&216.3s\\
     
        &Ratio&228.0&212.4&216.3\\
        \hline
                 \multirow{3}{*}{5000}&Exact&21243.6s&37717.0s&38199.5s\\
        
        &AAM&92.5s&198.7s&193.7s\\
       
        &Ratio&229.7&189.8&197.2\\
        \hline
    \end{tabularx}
    \caption{Comparison of the computational run-times between the exact method and AAM for different values of $\theta$ and $\rho$.}
    \label{tab:runtime}
\end{table}

\subsection{Optimal Design for Wide Field Telescope}\label{sec:widefov}

The on-axis angular resolution of a grazing incident X-ray telescope is very sensitive to figure errors in the optical mirrors with respect to the $WO_1$ shape. These figure errors mainly arise during the fabrication and assembly of the telescope. On the other hand, off-axis angular resolution is limited by the optical aberration associated with the $WO_1$. However, by modifying the optical design of $WO_1$ and adding a figure surface to it, the angular resolution variation with off-axis angle can be controlled to some extent. For a solar telescope with a very large field coverage up to $\pm 0.5^\circ$, the $WO_1$ design may provide very good on-axis angular resolution but with poor off-axis angular resolution. By adding this figure surface, in the form of a polynomial, to the $WO_1$, we can improve the off-axis angular resolution but at the cost of degrading the on-axis resolution.

To demonstrate DarsakX's capability in optimizing the design of a telescope with a wide FOV, we consider a telescope design suitable for a solar telescope with a wide FOV of $\pm 0.5^\circ$. The telescope is treated as a single shell with a focal length of 2 m and a radius of 20 cm. The length of both the primary and secondary mirrors is considered to be 3 cm. To optimize the angular resolution for the wide field in this telescope using DarsakX, we have applied fifth-degree polynomial figure surfaces to both the primary and secondary mirrors, with a base surface as $WO_1$. The equation for the figure surface, defined in the form of a polynomial, is as follows, where $x_0$ represents the focal length of the telescope.

\begin{equation}
    G_p(x)=\sum_{i=0}^{5}c_{i}(x-x_0)^i
\end{equation}

\begin{equation}
    G_h(x)=\sum_{i=0}^{5}c_{i}(x-x_0)^i
\end{equation}

\begin{table*}[h]
    \centering
    \begin{tabular}{ccccccc}
           \hline
         & $c_{0}$ & $c_{1}$  & $c_{2}$ & $c_{3}$ & $c_{4}$ & $c_{5}$\\
         \hline
       $G_p(x)$  & 0 & -6.1733e-05&  6.7033e-06 & -3.4419e-07 &  1.0736e-08 & -1.3058e-10 \\
         
        $G_h(x)$  & 0 & -5.2267e-05 & -3.8544e-06 &-6.8259e-08 & 4.6914e-10 &  1.9630e-11\\
          \hline
    \end{tabular}
    \caption{Polynomial coefficients}
    \label{tab:T16}
\end{table*}

\begin{figure*}[h]
  \begin{subfigure}{0.5\textwidth}
    \includegraphics[width=\linewidth]{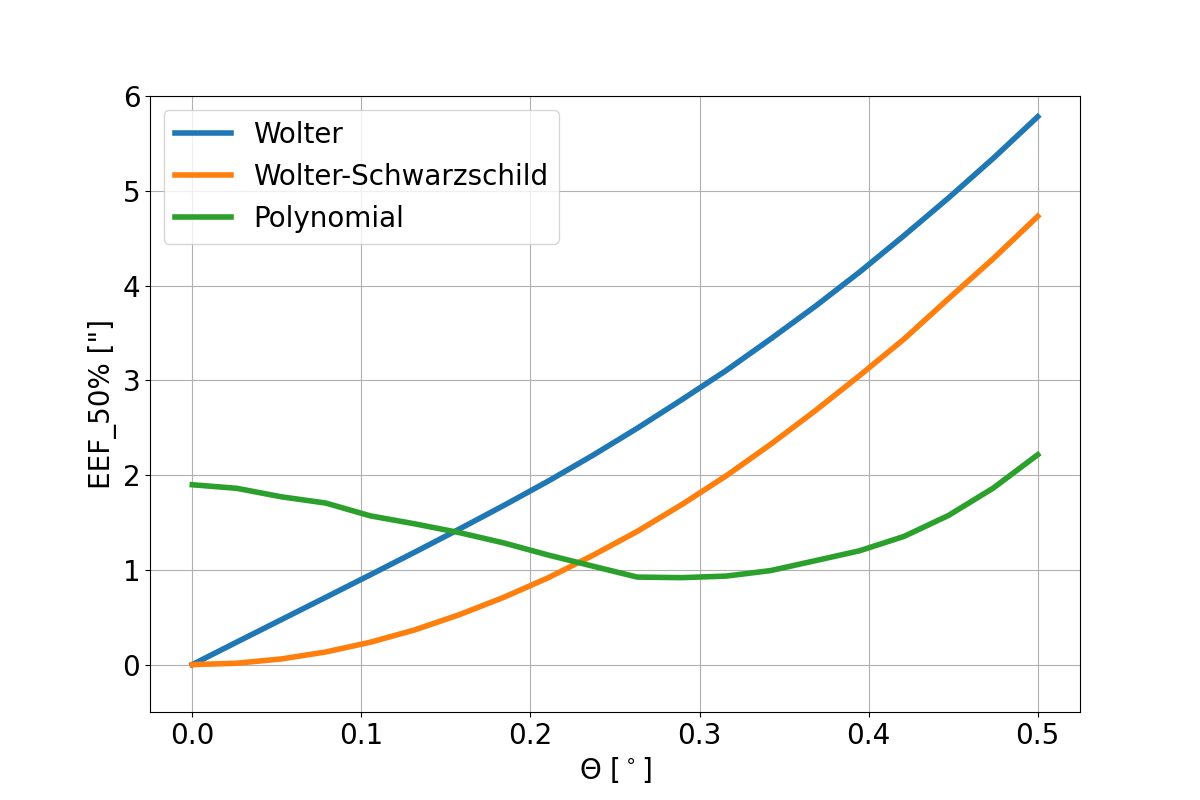}
    \caption{}
    \label{fig:widefov_a}
  \end{subfigure}%
  \begin{subfigure}{0.5\textwidth}
    \includegraphics[width=\linewidth]{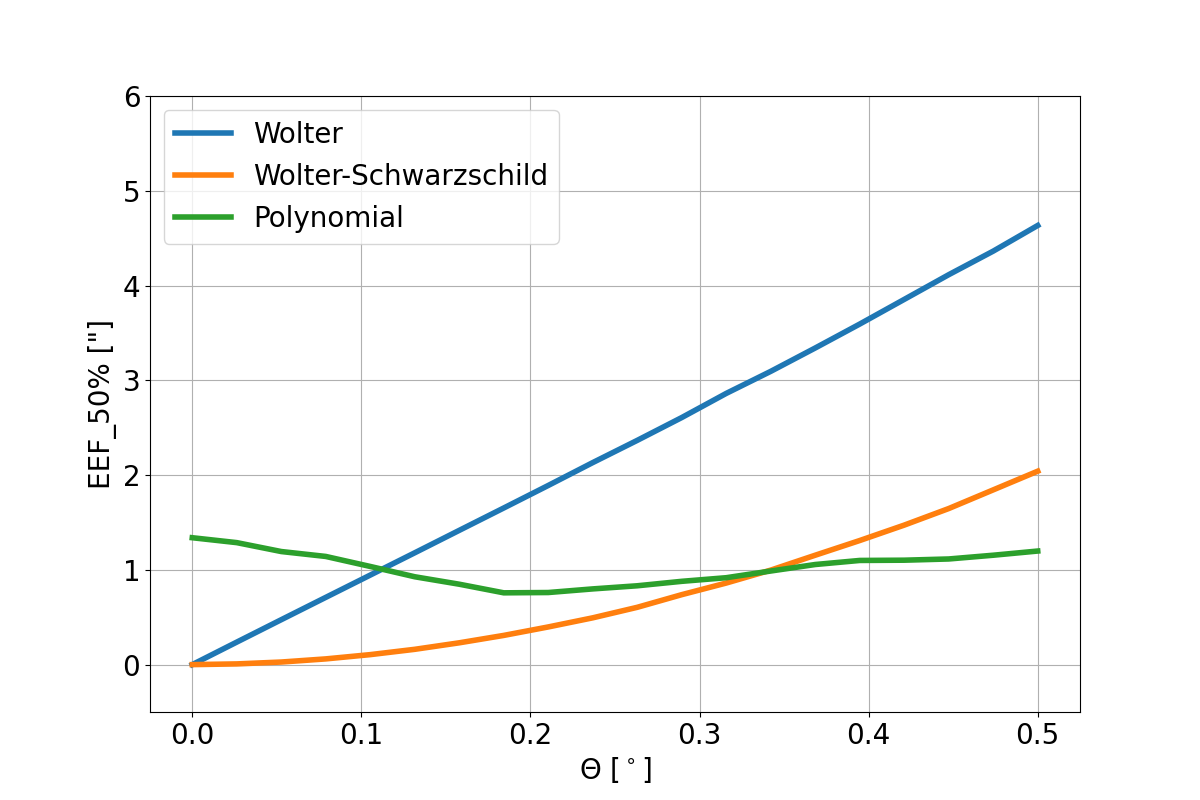}
    \caption{}
    \label{fig:widefov_b}
  \end{subfigure}
  \caption{Angular resolution variation with field angle $\theta$. (a) The focal detector is considered a plane. (b) The focal detector is considered as an optimized curved surface.}
  \label{fig:widefov}
\end{figure*}

\begin{figure*}[!h]
  \begin{subfigure}{0.5\textwidth}
    \includegraphics[width=\linewidth]{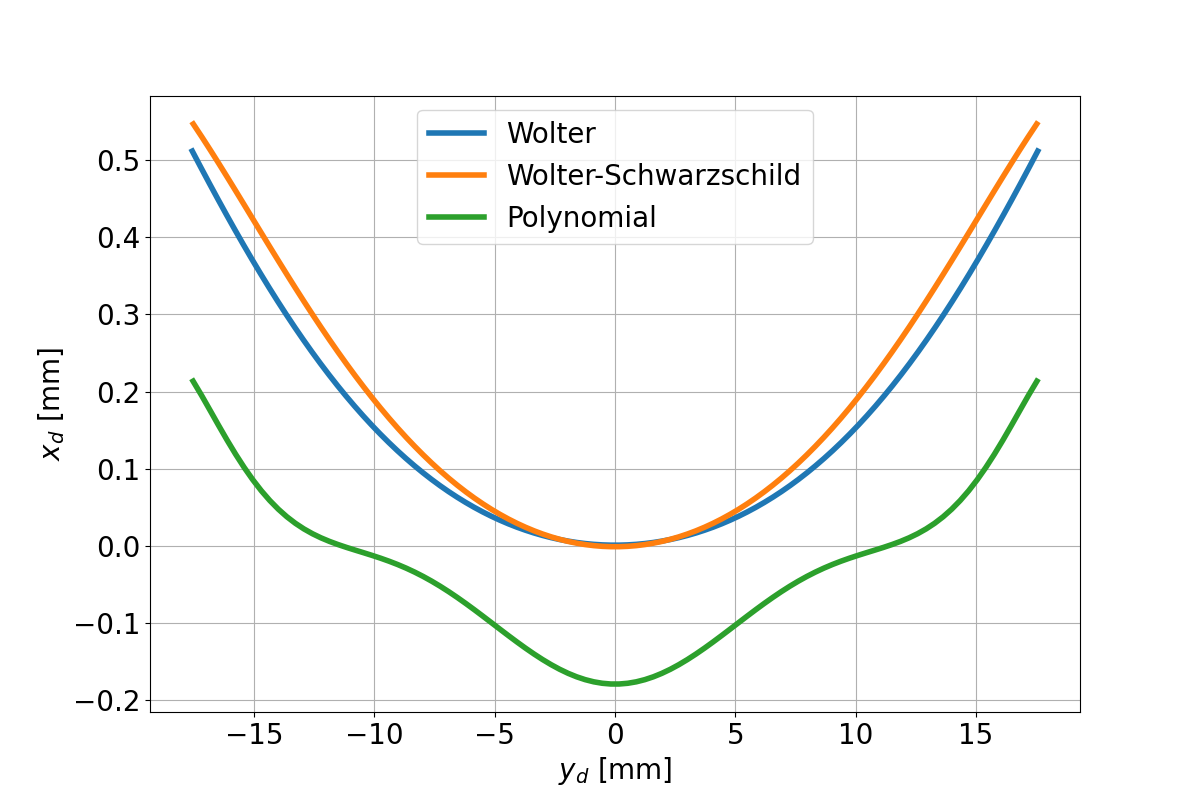}
    \caption{}
    \label{fig:widefov01_a}
  \end{subfigure}
  \begin{subfigure}{0.5\textwidth}
    \includegraphics[width=\linewidth]{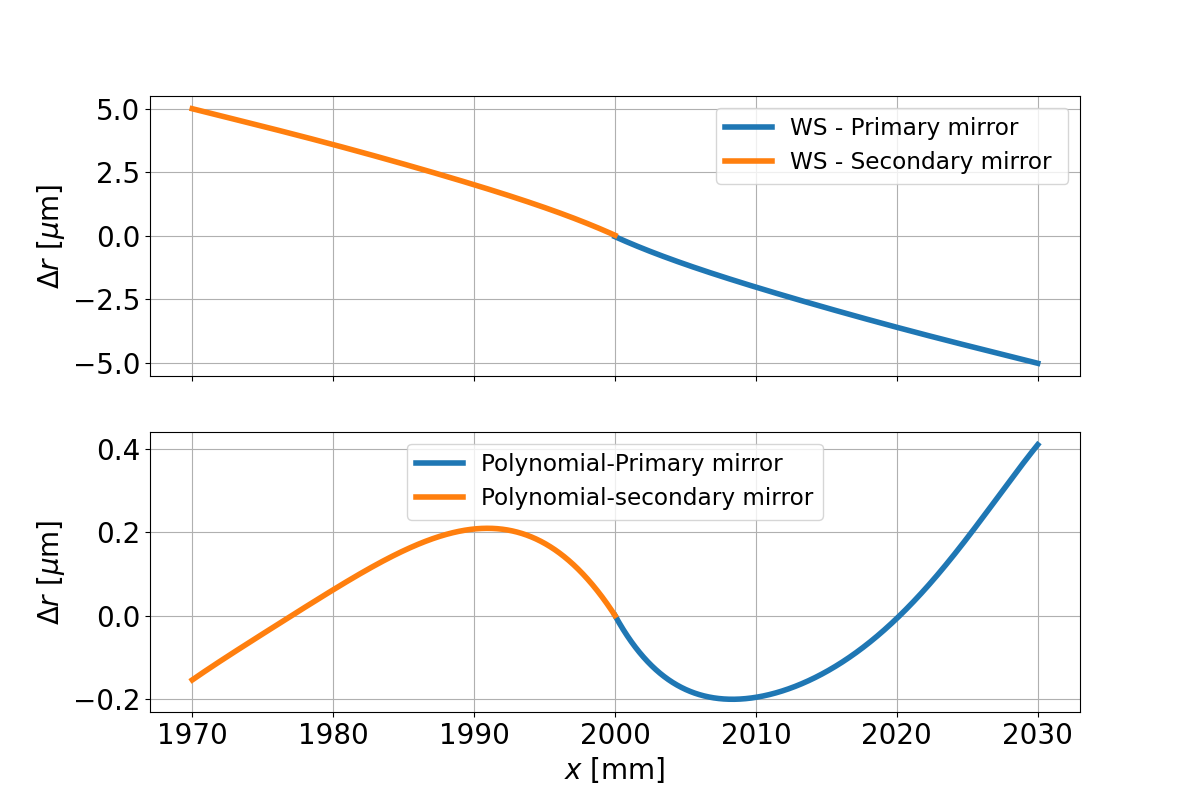}
    \caption{}
    \label{fig:widefov01_b}
  \end{subfigure}
  \caption{(a) The sectional view of the optimized focal surface. (b) The sectional view of the difference in the surface of polynomial type and $WS$ optics with respect to $WO_1$.}
  \label{fig:widefov01}
\end{figure*}

The polynomial coefficients obtained for the figure surface optimized to improve angular resolution are shown in Table \ref{tab:T16}. The optimized angular resolution demonstrates roughly uniform performance across the entire field of view, as shown in Figure \ref{fig:widefov}. When the focal surface is considered as a plane, polynomial-fitted optics provide better angular resolution compared to the $WO_1$ or $WS$, especially when the field angle is above $0.25^\circ$, as shown in Figure \ref{fig:widefov_a}. Even when the field angle is less than $0.25^\circ$, $EEF_{50\%}$ is within 2 arcsec. However, with the optimized focal surface, $WS$ provides better results, as shown in Figure \ref{fig:widefov_b}. The optimized focal surfaces are shown in Figure \ref{fig:widefov01_a} for all three optical designs. The polynomial figure surface is shown in Figure \ref{fig:widefov01_b} in the bottom section. This represents the radial difference in the $WO_1$ surface and the optical surface defined with a polynomial curve as the figure surface with the base surface as $WO_1$. The top section of Figure \ref{fig:widefov01_b} shows the radial difference between the $WS$ and $WO_1$ surfaces.

To obtain the optimal figure surface in the form of a polynomial, we iteratively altered the figure surface several times. Achieving this involves tracing rays through the telescope multiple times with different figure surfaces. Using the exact method for such optimizations is almost impossible as it would significantly increase the computational time. Instead, we obtained the optimized polynomial by iteratively tracing the rays multiple times using an AAM. After obtaining the optimized figure surface, the final results are plotted using the exact method.

\section{Summary}
DarsakX has been developed as a ray tracing tool designed to estimate the imaging performance of X-ray telescopes, with a specific focus on the future Indian X-ray observatory program. This tool is an extension of DarpanX, which was primarily developed to estimate the reflectivity of X-ray mirror foils. DarsakX is developed in such a way that it can be utilized to simulate any typical X-ray telescope configuration that follows the concept of two reflections, similar to Wolter-I optics. The tool has been designed with high flexibility, allowing users to control nearly all possible parameters of an X-ray telescope.

Ray tracing within DarsakX predominantly employs an analytical approach for fast and efficient simulations. However, when the need for numerical solutions arises, alternative options are available using AAM to expedite the simulation process. Comparative studies have demonstrated that these AAM yield highly acceptable results, especially when simulating realistic telescope configurations, in comparison to exact methods. Users have the choice between the exact method and AAM for their simulations.

DarsakX also enables the simulation of the effects of figure errors along the axial direction in the optics of X-ray telescopes. The software provides the option to select between Wolter-I and conical shapes as the base optics, considering the figure error. Validation of the DarsakX algorithm has been accomplished through comparative studies with the Chandra X-ray Observatory. The versatility of DarsakX is demonstrated by its capability to simulate telescopes with complex configurations, including multiple shells, exemplified by configurations similar to that of the XMM-Newton telescope.

The tool serves a dual purpose, as it can simulate telescope performance and aid in telescope design. It is expected that DarsakX will be employed iteratively to estimate the impact of figure errors on the imaging performance of telescopes with varying configurations. This iterative process will contribute to the enhancement of fabrication techniques, metrology for figure error measurements, and assembly methods. At present, figure errors in axial directions are only considered. In future developments, we plan to extend this to include azimuthal figure errors. Other future developments of DarsakX would include considering the effect of surface roughness in the mirror to account for X-ray scattering, enhancing 3D visualization, introducing a GUI-based user interface, and taking into account the impact of detectors and electronics on imaging performance.

\section*{Acknowledgements}
This work at the Physical Research Laboratory, Ahmedabad, is supported by the Department of Space, Government of India.

\appendix
\section{Ray Trace Through Single Shell}
\subsection{Base Surface as $WO_1$ }
\label{sec:Appendix}
\subsubsection{Ray Trace: Source to Paraboloid}
\label{subsec:SS1}
Let's assume the telescope receives parallel rays from a source at $\theta$ and $\phi=0$, as described in Figure \ref{fig:A1}. The direction of a ray, originated from source which incident on paraboloid section, has a direction as.

\begin{equation}  \label{eq:A10}
\hat{n}_{ip}=[-\cos\theta, -\sin\theta, 0]
\end{equation}

The rays originate from a source plane which has a normal in the direction of $\hat{n}_{ip}$, and this plane intersects $X$ axis at $[x_{s},0,0]$ where $x_{s}\geq(r_{0}\tan\theta+x_{0}+l_{p})$. A uniform mesh grid of the points ($[0,y_{s},z_{s}]$) is created on the source plane, in the coordinate frame of the source plane itself, which is rotated by an angle $\theta$ along $Z$ axis of telescope coordinate system. The grid spacing $l_{g}$ can be calculated as $l_{g}=1/\sqrt{\rho}$. Where $\rho $ is the ray density ($\mathrm{rays/cm^2}$). The coordinates of the grid points can be transformed from the source plane coordinate system to the telescope coordinate system as

\begin{equation}  \label{eq:A11}
\vec{R}_{s}=[x_{s}-y_{s}\sin\theta,y_{s}\cos\theta,z_{s}]
\end{equation}

The ray originated from $\vec{R}_{s}$, propagate in the direction of $\hat{n}_{ip}$, meet the paraboloid section at $\vec{R}_{p}$. Hence,

\begin{equation}  \label{eq:A12}
\frac{\vec{R}_{p}-\vec{R}_{s}}{\lVert\vec{R}_{p}-\vec{R}_{s}\rVert}= \hat{n}_{ip}
\end{equation}

Lets consider $\hat{n}_{ip}= [{n_{ip}}_{x}, {n_{ip}}_{y}, {n_{ip}}_{z}]$ and $\vec{R}_{s}= [{R_{s}}_{x}, {R_{s}}_{y}, {R_{s}}_{z}]$. Then

\begin{equation}  \label{eq:A13}
r_{p}^2=Ax_{p}^2+ Bx_{p} + C
\end{equation}
where,

\begin{equation}  \label{eq:A14}
A=\frac{{n_{ip}}_{y}^2}{{n_{ip}}_{x}^2}
\end{equation}

\begin{equation}  \label{eq:A15}
B=2\frac{{n_{ip}}_{y}}{{n_{ip}}_{x}}\left(R_{s_{y}}- R_{s_{x}}\frac{{n_{ip}}_{y}}{{n_{ip}}_{x}}\right)
\end{equation}

\begin{equation}  \label{eq:A16}
C=\left(R_{s_{y}}- R_{s_{x}}\frac{{n_{ip}}_{y}}{{n_{ip}}_{x}}\right)^2+R_{s_{z}}^2
\end{equation}  
 
The $x_{p}$ and $r_{p}$ can be solved numerically by comparing the Equation \ref{eq:A3} and Equation \ref{eq:A13}. However, if $G_{p}(x_{p})=0$, $x_{p}$ can be solved analytically by Equation \ref{eq:A17} and $r_{p}$ can be calculated through Equation \ref{eq:A13}.

\begin{equation} \label{eq:A17}
x_{p}=\frac{-B_0+\sqrt{B_0^2-4A_0C_0}}{2A_0}
\end{equation}

Where, $A_0=A$, $B_0=B-2p$ and $C_0=C-p^2-4pd\frac{e^2}{(e^2-1)}$. And $\phi_{p}$ can be calculated as.

\begin{equation}  \label{eq:A18}
\phi_{p}=\sin^{-1}\left(\frac{R_{s_{z}}}{r_{p}}\right)
\end{equation}

\subsubsection{Ray Trace: Paraboloid to Hyperboloid }
\label{subsec:SS2}
Since the incident location of the ray on the paraboloid section is known as $\vec{R}_{p}$ from previous section. A unit vector in the normal direction at a point  $\vec{R_{p}}$ on parabola is defined as $\hat{n}_{\perp p}$.

\begin{equation}  \label{eq:A19}
\hat{n}_{\perp p}=[\sin\beta_{p}, -\cos\phi_{p}\cos\beta_{p}, -\sin\phi_{p}\cos\beta_{p}]
\end{equation}

where

\begin{equation}  \label{eq:A20}
\tan\beta_{p}=\frac{dr_{p}}{dx}
\end{equation}

\begin{equation}  \label{eq:A21}
\frac{dr_{p}}{dx}=\frac{p}{(r_{p}-G_{p}(x))} + G'_{p}(x)
\end{equation}

The direction of reflected ray from paraboloid surface is defined as $\hat{n}_{rp}$. Now, after getting reflected from paraboloid, ray meets the hyperboloid surface at point $\vec{R}_{h}$. 

\begin{equation}  \label{eq:A22}
\hat{n}_{rp}=\hat{n}_{ip}-2\hat{n}_{\perp p}\left(\hat{n}_{\perp p}.\hat{n}_{ip}\right)
\end{equation}

Point $\vec{R}_{h}$ can be calculated by the following equation which state that ray reflected from the paraboloid surface moves along $\hat{n}_{rp}$ and finally meet hyperboloid surface at $\vec{R}_{h}$.

\begin{equation}  \label{eq:A23}
\frac{\vec{R}_{h}-\vec{R}_{p}}{\lVert\vec{R}_{h}-\vec{R}_{p}\rVert}= \hat{n}_{rp}
\end{equation}

Lets consider  $\hat{n}_{rp}= [{n_{rp}}_{x}, {n_{rp}}_{y}, {n_{rp}}_{z}]$. By solving the Equation \ref{eq:A23} we can get the following equation.

\begin{equation}  \label{eq:A24}
r_{h}^2=A_{1}x_{h}^2+ B_{1}x_{h} + C_{1}
\end{equation}

Where

\begin{equation}  \label{eq:A25}
A_1=\frac{\left({n_{rp}}_{y}^2 + {n_{rp}}_{z}^2\right)}{{n_{rp}}_{x}^2}
\end{equation}

\begin{equation}  \label{eq:A26}
B_1= 2\left( r_{p}\frac{\left({n_{rp}}_{y}\cos\phi_{p}+{n_{rp}}_{z}\sin\phi_{p}\right)}{{n_{rp}}_{x}} -x_{p} \frac{\left({n_{rp}}_{y}^2 + {n_{rp}}_{z}^2\right)}{{n_{rp}}_{x}^2}  \right)
\end{equation}

\begin{equation}  \label{eq:A27}
C_1= r_{p}^2+ x_{p}^2 \frac{\left({n_{rp}}_{y}^2 + {n_{rp}}_{z}^2\right)}{{n_{rp}}_{x}^2}
- 2 x_{p} r_{p}\frac{\left({n_{rp}}_{y}\cos\phi_{p}+{n_{rp}}_{z}\sin\phi_{p}\right)}{{n_{rp}}_{x}}
\end{equation} 

The $x_{h}$ and $r_{h}$ can be solved numerically by comparing the Equation \ref{eq:A4} and Equation \ref{eq:A24}. However, if $G_{h}(x_{h})=0$, $x_{h}$ can be solved analytically by Equation \ref{eq:A28} and $r_{h}$ can be calculated through Equation \ref{eq:A24}.

\begin{equation} \label{eq:A28}
x_{h}=\frac{-B_{10}+\sqrt{B_{10}^2-4A_{10}C_{10}}}{2A_{10}}
\end{equation}

Where, $A_{10}=A_1-(e^2-1)$, $B_{10}=B_1-2de^2$ and $C_{10}=C_1-e^2d^2$. And $\phi_{h}$ can be calculates as.

\begin{equation}  \label{eq:A29}
\phi_{h}=\sin^{-1} \left(\frac{{n_{rp}}_{z}}{{n_{rp}}_{x}}\frac{\left( x_{h}-x_{p}\right)}{r_{h}} + \frac{r_{p}\sin\phi_{p}}{r_{h}}\right)
\end{equation}

\subsubsection{Ray Trace: Hyperboloid to Detector }
\label{subsec:SS3}
A unit vector in the normal direction at a point  $\vec{R_{h}}$ on hyperboloid surface is defined as $\hat{n}_{\perp h}$.

\begin{equation}  \label{eq:A30}
\hat{n}_{\perp h}=[\sin\beta_{h}, -\cos\phi_{h}\cos\beta_{h}, -\sin\phi_{h}\cos\beta_{h}]
\end{equation}

Where.

\begin{equation}  \label{eq:A31}
\tan\beta_{h}=\frac{dr_{h}}{dx}
\end{equation}

\begin{equation}  \label{eq:A32}
\frac{dr_{h}}{dx}= \frac{e^2(x+p)-x}{(r_{h}-G_{h}(x))} + G'_{h}(x)
\end{equation}

The direction of reflected ray from hyperboloid surface is defined as $\hat{n}_{rh}$. The ray reflected from hyperboloid surface at $\vec{R}_{h}$ moves along the direction of $\hat{n}_{rh}$ until it meets the detector plane at $\vec{R}_{d}$.

\begin{equation}  \label{eq:A33}
\hat{n}_{rh}=\hat{n}_{rp}-2\hat{n}_{\perp h}\left(\hat{n}_{\perp h}.\hat{n}_{rp}\right)
\end{equation}

\begin{equation}  \label{eq:A34}
\vec{R}_{d}=[x_d, y_{d}, z_{d}]
\end{equation}

The detector plane is defined as a plane which has normal along $X$ axis and it is placed at $x=x_d$. The point $\vec{R}_{d}$ on detector plane can be calculate by solving the following following equation.

\begin{equation}  \label{eq:A35}
\frac{\vec{R}_{d}-\vec{R}_{h}}{\lVert{\vec{R}_{d}-\vec{R}_{h}}\rVert}= \hat{n}_{rh}
\end{equation}
If we define  $\hat{n}_{rh}= [{n_{rh}}_{x}, {n_{rh}}_{y}, {n_{rh}}_{z}]$, then $y_d$ and $z_d$ can be solved as.

\begin{equation}  \label{eq:A36}
y_{d}=r_{h}\cos\phi_{h}+\left( x_d-x_{h}\right)\frac{{n_{rh}}_{y}}{{n_{rh}}_{x}}
\end{equation}

\begin{equation}  \label{eq:A37}
z_{d}=r_{h}\sin\phi_{h}+\left( x_d-x_{h}\right)\frac{{n_{rh}}_{z}}{{n_{rh}}_{x}}
\end{equation} 

\subsection{Base Surface as $CO$ }
\label{sec:Appendix_CO}

All the steps described to trace a ray for $WO_1$ case in \ref{sec:Appendix} will be applicable to conical approximation as well except the few changes that are listed. \\

The equation describing the base surface defined by Equation \ref{eq:A3} and \ref{eq:A4}, will now be substituted with Equations \ref{eq:A46} and \ref{eq:A47}. Additionally, when solving the intersection of the ray with primary mirror, if we solve the case when $G_{p}(x_{p})=0$, $A_0$, $B_0$ and $C_0$ will be changed as $A_{0}=A-({\tan\beta_{p0}})^2$, $B_{0}=B-2\tan\beta_{p0}\left(r_0-x_0\tan\beta_{p0}\right)$ and $C_{0}=C-\left(r_0-x_0\tan\beta_{p0}\right)^2$.
\\

Similarly if we solve the case for $G_{h}(x_{h})=0$, $A_{10}$, $B_{10}$ and $C_{10}$ will be changed as $A_{10}=A_1-({\tan\beta_{h0}})^2$, $B_{10}=B_1-2\tan\beta_{h0}\left(r_0-x_0\tan\beta_{h0}\right)$ and $C_{10}=C_1-\left(r_0-x_0\tan\beta_{h0}\right)^2$.
\\

The slopes at the points where the ray strikes the primary and secondary mirrors, as represented by Equations \ref{eq:A21} and \ref{eq:A32} in $WO_1$, will be changed as follows:

\begin{equation}  \label{eq:A49}
\frac{dr_{p}}{dx}=\tan\beta_{p0} + G'_{p}(x)
\end{equation}

\begin{equation}  \label{eq:A50}
\frac{dr_{h}}{dx}=\tan\beta_{h0} + G'_{h}(x)
\end{equation}

Rest everything will be the same as $WO_1$.

\subsection{Ray Trace by AAM}
\label{sec:Appendix_3}

\begin{figure}[H]
\begin{center}
  \includegraphics[width=1\linewidth]{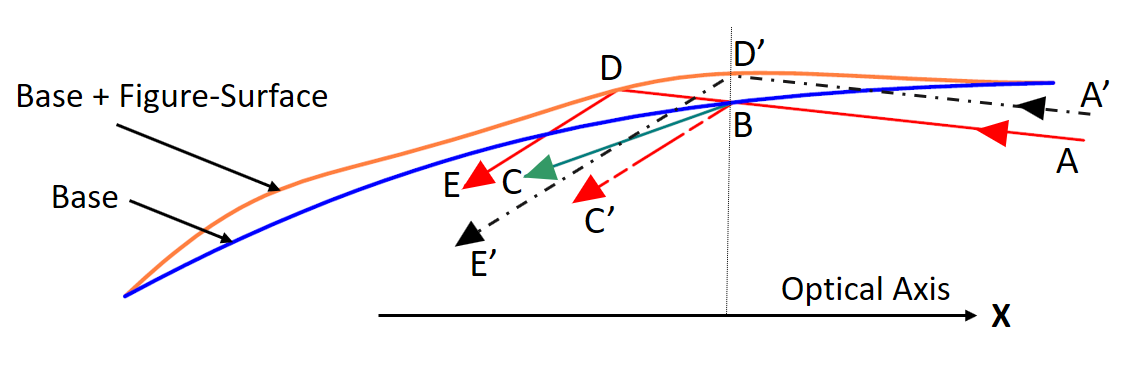}
  \caption{Ray trajectory diagram `with' and `without' figure-surface.}
  \label{fig:A1_1}
\end{center}
\end{figure}

Figure \ref{fig:A1_1} illustrates a ray tracing diagram for a single mirror within a shell, whether a primary or a secondary one. This diagram demonstrates how the AAM operates. The base surface refers to the primary or secondary mirror surface in $WO_1$ or $CO$ configuration. The term `Base + figure surface' defines the base surface along with the figure surface.

Ray \textit{ABC} represents the ideal path when a ray undergoes reflection from a mirror with only a base surface. If the same ray reflects from a surface with a non-zero figure surface in addition to the base surface, it follows the path \textit{ADE}. While the ideal path \textit{ABC} can be solved analytically, tracing the actual path \textit{ADE} requires a numerical solution because the coordinates of point \textit{D} cannot always be obtained analytically due to the figure surface, which can be non-linear.

Therefore, in the AAM, we consider that the ray reflects from the ideal surface (from point \textit{B}) instead of the actual surface (from point \textit{D}). However, the direction of reflection remains actual (as \textit{DE}) rather than ideal (as \textit{BC}). Thus, in AAM, we depict the ray path as \textit{ABC\'}, where \textit{BC\'} is parallel to \textit{DE}. This indicates that the ray reflects from the ideal surface but in the actual direction. 

Since in AAM, the coordinates of point \textit{D} are unknown, direct estimation of the direction \textit{DE} is not feasible. Hence in AAM, we obtain \textit{D\'{}E\'} instead of \textit{DE} as both are nearly parallel. Here, \textit{D\'} and \textit{B} share the same axial position and \textit{A\'{}D\'} is considered to be parallel to \textit{AB}. Given that the figure surface is on the order of sub-micron levels, the points \textit{D}, \textit{D\'}, and \textit{B} are very close, operating at the micron scale. Consequently, the final deviation in the ray's position on the focal plane is more influenced by the change in the ray's direction (either reflected in the direction of \textit{DE} or \textit{BC}), rather than the position of the reflected point (either \textit{D} or \textit{B}).

In AAM, the ray tracing methodology will remain the same as described in \ref{sec:Appendix} and \ref{sec:Appendix_CO}, with a few specified changes outlined below.

When determining the intersection point of the ray with the primary and secondary mirrors, the figure surfaces ($G_p(x)=0$ and $G_h(x)=0$) will be considered as zero. Consequently, the axial positions of the ray's intersections with the primary and secondary mirrors will be determined using Equations \ref{eq:A17} and \ref{eq:A28}, respectively.

However, in calculating the slope of the primary and secondary curves at the points where the ray intersects them, neither the figure surfaces ($G_{p}(x)\neq0$, $G_{h}(x)\neq0$) nor their derivatives ($G'_{p}(x)\neq0$, $G'_{h}(x)\neq0$) will be considered as zero. Therefore, the slopes of the curves will be evaluated using Equations \ref{eq:A21} and \ref{eq:A32} in the $WO_1$ case, and Equations \ref{eq:A49} and \ref{eq:A50} in the $CO$ case.

\bibliographystyle{elsarticle-harv}
\bibliography{references}

\end{document}